\documentclass[traditabstract]{aa}
\usepackage{graphicx}
\usepackage{txfonts}
\usepackage{natbib}
\usepackage{enumitem}
\bibpunct{(}{)}{;}{a}{}{,}

 \newcommand{\mics}{$\mu$m}
 
  \newcommand{\msun}{M$_{\odot}$}
  \newcommand{\lsun}{L$_{\odot}$}
 \newcommand{\lfir}{$L _{60}$}
 \newcommand{\lhard}{$L _{X}$}
 \newcommand{\smass}{$M _{*}$}
 \newcommand{\reff}{$R _{e}$} 
 \newcommand{\ergs}{erg s$^{-1}$}
 \newcommand{\kms}{km s$^{-1}$} 
 \newcommand{\sersic}{S\'ersic}

\begin{document}

\title{The host galaxies of X-ray selected Active Galactic Nuclei to $z=2.5$: Structure, star-formation and their relationships from CANDELS and Herschel/PACS\thanks{Herschel is an ESA space observatory with science instruments provided by European-led Principal Investigator consortia and with important participation from NASA.}}

\author{
D. J. Rosario\inst{1}
\and
D. H. McIntosh\inst{2}
\and
A. van der Wel\inst{3}
\and
J. Kartaltepe\inst{4}
\and
P. Lang\inst{1}
\and
P. Santini\inst{5}
\and
S. Wuyts\inst{1}
\and
D. Lutz\inst{1}
\and
M. Rafelski\inst{6}
\and
C. Villforth\inst{22, 39}
\and 
D. M. Alexander\inst{7}
\and
F. E. Bauer\inst{8,23}
\and
E. F. Bell\inst{9}
\and
S. Berta\inst{1}
\and
W. N. Brandt\inst{10}
\and
C. J. Conselice\inst{11}
\and
A. Dekel\inst{12}
\and
S. M. Faber\inst{13}
\and
H. C. Ferguson\inst{14}
\and 
R. Genzel\inst{1}
\and
N. A. Grogin\inst{14}
\and 
D. D. Kocevski\inst{15}
\and
A. M. Koekemoer\inst{14}
\and 
D. C. Koo\inst{13}
\and
J. M. Lotz\inst{14}
\and
B. Magnelli\inst{16}
\and
R. Maiolino\inst{17,18}
\and
M. Mozena\inst{13}
\and
J. R. Mullaney\inst{19}
\and
C. J. Papovich\inst{20}
\and
P. Popesso\inst{21}
\and
L. J. Tacconi\inst{1}
\and
J. R. Trump\inst{10}
\and
S. Avadhuta\inst{14}
\and
R. Bassett\inst{24}
\and
A. Bell\inst{9}
\and
M. Bernyk\inst{24}
\and
F. Bournaud\inst{25}
\and
P. Cassata\inst{26}
\and
E. Cheung\inst{13}
\and
D. Croton\inst{24}
\and
J. Donley\inst{27}
\and
L. DeGroot\inst{28} 
\and
J. Guedes\inst{29}
\and
N. Hathi\inst{30}
\and
J. Herrington\inst{9}
\and
M. Hilton\inst{31}
\and
K. Lai\inst{13}
\and
C. Lani\inst{27}
\and
M. Martig\inst{24}
\and
E. McGrath\inst{32}
\and
S. Mutch\inst{24}
\and
A. Mortlock\inst{11}
\and
C. McPartland\inst{33}
\and
E. O'Leary\inst{3,34}
\and
M. Peth\inst{35}
\and
A. Pillepich\inst{36}
\and
G. Poole\inst{24}
\and
D. Snyder\inst{13}
\and
A. Straughn\inst{37}
\and
O. Telford\inst{38}
\and
C. Tonini\inst{24}
\and
P. Wandro\inst{13}
}

\offprints{D. Rosario \email{rosario@mpe.mpg.de}}

\institute{Max-Planck-Institut f\"{u}r Extraterrestrische Physik (MPE), Postfach 1312, 85741 Garching, Germany   
\and Department of Physics \& Astronomy, University of Missouri-Kansas City, 5110 Rockhill Rd., Kansas City, MO 64110, USA  
\and Max-Planck Institut f\"{u}r Astronomie, K\"{o}nigstuhl 17, D-69117, Heidelberg, Germany  
\and National Optical Astronomy Observatory, 950 N. Cherry Ave., Tucson, AZ, 85719, USA  
\and INAF - Osservatorio Astronomico di Roma, via di Frascati 33, 00040 Monte Porzio Catone, Italy  
\and Infrared Processing and Analysis Center, California Institute of Technology, Pasadena, CA, USA  
\and Department of Physics, Durham University, South Road, Durham DH1 3LE, UK  
\and Instituto de Astrof\'{\i}sica, Facultad de F\'{i}sica, Pontificia Universidad Cat\`{o}lica de Chile, 306, Santiago 22, Chile  
\and Department of Astronomy, University of Michigan, 500 Church St., Ann Arbor, MI 48109  
\and Department of Astronomy \& Astrophysics, The Pennsylvania State University, University Park, Pennsylvania, PA 16802 USA   
\and The School of Physics and Astronomy, University of Nottingham, Nottingham, UK  
\and Racah Institute of Physics, The Hebrew University, Jerusalem, Israel  
\and University of California Observatories/Lick Observatory, University of California, Santa Cruz, CA 95064, USA  
\and Space Telescope Science Institute, 3700 San Martin Drive, Baltimore, MD 21218  
\and  Department of Physics and Astronomy, University of Kentucky, Lexington KY 40506-0055, USA   
\and Argelander-Institut f\"{u}r Astronomie, Auf dem H\"{u}gel 71, D-53121 Bonn  
\and Kavli Institute for Cosmology, University of Cambridge, Madingley Road, Cambridge CB3 OHA, UK  
\and Cavendish Laboratory, University of Cambridge, 19 JJ Thomson Avenue, Cambridge, CB3 OHE, UK   
\and Department of Physics and Astronomy, University of Sheffield, Hounsfield Road, Sheffield S3 7RH, UK 
\and Department of Physics and Astronomy, Texas A\&M University, College Station, TX, USA 
\and Exzellenzcluster Universe, Technische Universit\"at M\"unchen, Boltzmannstrasse 2, D-85748 Garching, Germany 
\and SUPA, School of Physics and Astronomy, University of St. Andrews, North Haugh, St. Andrews, Fife KY16 9SS, UK
\and Space Science Institute, 4750 Walnut Street, Suite 205, Boulder, Colorado 80301 
\and Centre for Astrophysics \& Supercomputing, Swinburne University of Technology, P.O. Box 218, Hawthorn, VIC 3122, Australia 
\and Laboratoire AIM-Paris-Saclay, CEA/DSM/Irfu - CNRS - Universit\'e Paris Diderot, CE-Saclay, F-91191 Gif-sur-Yvette, France 
\and Aix Marseille Universit\'e, CNRS, LAM (Laboratoire d?Astrophysique de Marseille) UMR 7326, 13388, Marseille, France 
\and Los Alamos National Laboratory, Los Alamos NM 
\and Department of Physics and Astronomy, University of California, Riverside, CA 92521, USA 
\and Institute for Astronomy, ETH Zurich, Wolgang-Pauli-Strasse 27, 8093 Zurich, Switzerland 
\and Aix Marseille Universit\'{e}, CNRS, LAM (Laboratoire d'Astrophysique de Marseille) UMR 7326, 13388, Marseille, France
\and School of Mathematics, Statistics \& Computer Science, University of KwaZulu-Natal, Durban 4041, South Africa 
\and Department of Physics and Astronomy, Colby College, Waterville, ME 04901, USA 
\and Institute for Astronomy, University of Hawaii, 2680 Woodlawn Drive, Honolulu, HI 96822 
\and Department of Physics and Astronomy, Macalester College, 1600 Grand Avenue, Saint Paul, MN 55105, USA
\and Department of Physics and Astronomy, Johns Hopkins University, 3400 North Charles Street, Baltimore, MD 21218, USA
\and Harvard-Smithsonian Center for Astrophysics, 60 Garden Street, Cambridge, MA 02138, USA
\and NASA Goddard Space Flight Center
\and Astronomy Department,  3910 15th Ave NE, University of Washington, Seattle, WA 98195, USA
\and Department of Astronomy, University of Florida, 211 Bryant Space Science Center, Gainesville, FL 32611-2055, USA 
}

\titlerunning{Structure and SFR of AGN in CANDELS/CDF}

 \abstract{We study the relationship between the structure and star-formation rate (SFR) of X-ray selected low and moderate luminosity
 active galactic nuclei (AGNs) in the two Chandra Deep Fields, using Hubble Space Telescope imaging from the
 Cosmic Assembly Near Infrared Extragalactic Legacy Survey (CANDELS) and deep far-infrared maps from the
 PEP+GOODS-Herschel survey.  We derive detailed distributions of structural parameters and 
 FIR luminosities from carefully constructed control samples of galaxies, which we then compare to those of the AGNs.
 At $z\sim1$, AGNs show slightly diskier light profiles than massive inactive (non-AGN) galaxies, as well as 
 modestly higher levels of gross galaxy disturbance (as measured by visual signatures of interactions and clumpy structure). 
 In contrast, at $z\sim2$, AGNs show similar levels of galaxy disturbance as inactive galaxies, but display a red central light enhancement, which
 may arise due to a more pronounced bulge in AGN hosts or due to extinguished nuclear light. We undertake a number of tests
 of both these alternatives, but our results do not strongly favour one interpretation over the other.
 The mean SFR and its distribution among AGNs and inactive galaxies are similar at $z>1.5$. 
 At $z<1$, however, clear and significant enhancements are seen in the SFRs of AGNs with
 bulge-dominated light profiles. These trends suggest an evolution in the relation between nuclear activity and host
 properties with redshift, towards a minor role for mergers and interactions at $z>1.5$.
  }

\maketitle

\section{Introduction}

The study of galaxy structure (or morphology, used interchangeably in this work) 
has provided important insights into galaxy formation and evolution and is a key observable in constraints on theoretical models. 
It is conclusively linked to other important galaxy properties such as mass, baryonic content, star-formation history, 
interaction state and environment \citep[e.g.,][]{dressler80,roberts94,kennicutt98,strateva01,wuyts11}. In this sense,
structure is a sensitive measure of the history of galaxy growth and can be used to constrain
galaxy evolution models. Another important tracer of the evolutionary state of a galaxy is the star-formation rate (SFR).
This is sensitive to the gas content and infall onto a galaxy, as well as the processes that
inject energy and momentum into the ISM and regulate the overall efficiency of cold gas fragmentation and the formation of new stars. 

Active Galactic Nuclei (AGNs) are found in a subset of galaxies where a central supermassive black hole (SMBH)
is accreting material at a sufficient level to be detectable using many characteristic tracers of high-energy activity, such
as strong X-ray, non-thermal radio, hot dust or high excitation line emission. There are many theories
as to what triggers and fuels SMBH accretion (e.g., \citealp{hopkins05a,fanidakis12,hirschmann12}; see \citealp{alexander12} for a
contemporary discussion). These may be broadly divided into secular processes 
(bars and spirals, possibly minor mergers) that take several galaxy dynamical timescales ($t_{dyn}$)
to bring gas to the nucleus from galaxy scales, or ``violent" processes (harassment,
major mergers, violent disk instability), which operate through varying, typically external, torques
that change on $t_{dyn}$, and are expected to circumvent secular inflow.
Since $t_{dyn}$ is the characteristic duration over which a galaxy responds to a violent disturbance
and relaxes into a new state of dynamical equilibrium, the relationship between galaxy structure and the occurrence
or strength of nuclear activity is a vital indicator of the relative importance of violent and secular AGN fueling mechanisms.

Star-formation  is also affected differently by these two classes of processes. Stable galactic gas
disks form stars with a relatively low efficiency, which current studies suggest
is modulated by the interplay of dense gas and stellar feedback \citep{kim13,hopkins13b}. Violent processes lead to the strong inflow of gas into the
centers of galactic potential wells \citep{sanders88} or into dense small-scale clumps \citep{dekel09}, both of which result in star-forming environments
with an enhanced efficiency of star-formation \citep{genzel10, daddi10b}. Observationally, 
star-forming galaxies show a clear correlation between SFR and stellar mass (\smass)
\citep{noeske07, elbaz07, daddi07, santini09, rodighiero11}, sometimes called the star-formation (SF) sequence. 
The existence of this sequence implies that secular processes govern most
star-forming galaxies and, indeed, structural studies of galaxies across much of cosmic time show that galaxies
on the SF sequence are disk-dominated, indicative of low levels of dynamical disturbance \citep{forsterschreiber09,wuyts11}.
Starbursts are defined to have an elevated SFR, placing them above the SF sequence \citep{rodighiero11,sargent12}, 
and are structurally consistent with being recently disturbed \citep{wuyts11, kartaltepe12}. This association between 
structure and SF patterns highlights alternate pathways for violent
and secular processes, in terms of the connection between SF and the morphological change in galaxies. 

The morphologies of AGN hosts have been extensively studied in various different ways, across a wide range
of redshifts. 
In general, these studies have
either used analytic measures of structure, such as light profile fitting, pixel distribution statistics 
or moment measures \citep{grogin03, grogin05, pierce07, gabor09, simmons11,schawinski11,bohm13,villforth14} 
or visual classification using a number of special-purpose schemes \citep{schawinski07,schawinski10,
cisternas11, kocevski12}. Some studies of more luminous optically-unobscured AGNs have also
been undertaken, involving the careful subtraction of nuclear point sources to reveal the host galaxy
more clearly \citep[e.g.,][]{bahcall97, mcleod01, dunlop03, jahnke04, sanchez04,guyon06, veilleux09}. 
AGN hosts span a range of morphologies and levels of galaxy disturbance, with a sizeable
fraction found in massive galaxies with substantial disks and low levels of merger activity. The most luminous
systems tend to be in more bulge-dominated or early-type hosts.

In parallel, the SF properties of AGN hosts is topic of much current work. Various tracers of the SFR
have been applied to AGNs, though traditional ones such as emission lines, UV continuum or optical spectral features
can be strongly contaminated by nuclear light, especially among luminous Type I AGNs. Despite the limited sensitivity
of current datasets, far-infrared (FIR) wavelengths ($>50$ \mics)
offer the best discrimination between emission from star-formation heated dust and AGN-heated dust, since the latter
almost always exhibits a warmer distribution of temperatures, peaking in luminosity at $\sim20$ \mics\ with a sharp
drop off to the FIR \citep{netzer07, mullaney11, rosario12}. Herschel-based SFRs of active galaxies 
across a range of luminosities have been shown to agree reasonably well with those of massive inactive (i.e., non-AGN)
star-forming galaxies \citep{mullaney12, santini12, rovilos12, rosario13b, rosario13c}, but there is a strong tendency
for X-ray bright AGNs to be found in SF hosts \citep{rosario13b}. Therefore, AGN hosts are primarily 
massive galaxies forming stars at a normal rate given their stellar content. 

In this work, we present an exploration into connections between the SFR and structure of AGN hosts 
at intermediate to high redshifts ($z=[0.5,2.5]$), towards constraining the degree to which AGNs exhibit
the relationships known among inactive galaxies (i.e., those without substantial on-going SMBH accretion).
The primary goal of this paper is to lay the groundwork for a more extensive and detailed analysis of such relationships
using forthcoming datasets spanning a number of important extragalactic fields. Despite this, our work
is already unique in its approach of testing SFR-structure trends in systems with significant AGN activity across
an important range in redshifts.
Our study builds on the best available deep extragalactic data at X-ray, optical, NIR and FIR wavelengths, and
leverages the substantial efforts of the extragalactic community towards building coherent datasets in the GOODS fields.
Datasets, selections and methods are introduced in Section 2. 
We discuss specific structural patterns relevant to the galaxy population at these redshifts and compare them to
the AGNs, revealing any differences between these two populations (Section 3). In Section 4, we investigate trends
between SFR and structure, again comparing the AGNs to the galaxy population. Finally, in Section 5, we discuss our
results in the context of earlier work and weigh them qualitatively against the predictions of key AGN triggering scenarios.

We adopt a standard $\Lambda$-CDM Concordance cosmology, with $H_{0} = 70$ \kms~Mpc$^{-1}$ and $\Omega_{\Lambda}=0.7$. 
Stellar masses in this study are estimated assuming a Salpeter Initial Mass Function. 

\section{Sample Selection, Datasets and Methods}

In this work, we concentrate on X-ray selected AGN, since luminous X-ray emission is an 
unambiguous hallmark of SMBH accretion activity. However, studies of the AGN population and the X-ray background
suggest that X-ray selection systematically misses the most obscured AGNs. Our conclusions from this
work are strictly applicable only to such AGNs as are traced by X-ray selection. This caveat must be borne in mind
when setting this work into the overall context of AGN demographics.

\subsection{CANDELS HST imaging}

A panchromatic spatially resolved view of substantial numbers of galaxies in the distant Universe is now available
through the Cosmic Near-Infrared Deep Extragalactic Legacy Survey (CANDELS), which consists of HST ACS and WFC3
imaging of five separate deep extragalactic fields in multiple broad photometric bands spanning the observed
UV to the near-IR. Full details of the design, data reduction and available formats of the survey are discussed in 
\citet{grogin11} and \citet{koekemoer11}. 
The great leap forward for galaxy structure work made possible by CANDELS comes from deep near-infrared
imaging from the IR channel of the WFC3 camera. The reddest CANDELS band is F160W, which approximately corresponds
to the H-band at 1.6 \mics. For galaxies at $z=2.5$, F160W lies completely redward of rest-frame 4000 \AA. At these wavelengths,
the continuum light from galaxies is representative of the total stellar content of the systems, and is significantly
less biased by rest-frame ultra-violet light from young stars that can completely dominate in the observed optical and UV wavelengths. Earlier
studies of galaxy structure over large galaxy samples at HST resolution were restricted to observed wavelength $<1$ \mics.
With its capacity to capture the true stellar distribution of statistically complete samples of galaxies, CANDELS has greatly expanded
the abilities of galaxy structural analysis at $z\sim2$.

We measure structure from the entire CANDELS mosaics 
in the two fields associated with the Great Observatories Origins Deep Survey \citep[GOODS][]{giavalisco04}. 
The CANDELS mosaic in GOODS-S consists of an ultradeep region (Hubble Ultra-Deep field; $2'\times2\farcm3$) enclosed by 
a deep region (CANDELS-DEEP; $6\farcm8\times10'$), both of which are enclosed by a larger shallower region
(CANDELS-Wide; $10'\times16'$). The CANDELS mosaic in GOODS-N is similar in size to that in GOODS-S, with Deep and Wide
subregions but without an ultradeep section.

The HST bands cover different rest-frame wavelengths across the redshifts considered
in this work. Therefore, it is important to keep in mind that that when we discuss or highlight 
evolution in structural properties obtained from our own analysis, we specifically
mean apparent evolution which may be subject to morphological K-corrections. The main conclusions of the paper are based on
structural and SFR comparisons between galaxy populations at similar redshifts, which are not subject to K-corrections.
As a rough gauge for the reader, F160W (H-band) covers a rest-frame band at $\approx 8000$\AA\ at $z=1$ and 
$\approx 5200$\AA\ at $z=2$. Two other HST optical bands relevant for appreciating the following analysis are
 F606W (V-band; $\approx 3000$\AA\ at $z=1$ and $\approx 2000$\AA\ at $z=2$) and 
 F850LP  (z-band; $\approx 4500$\AA\ at $z=1$ and $\approx 3000$\AA\ at $z=2$). 

\begin{figure*}[t]
\includegraphics[width=\textwidth]{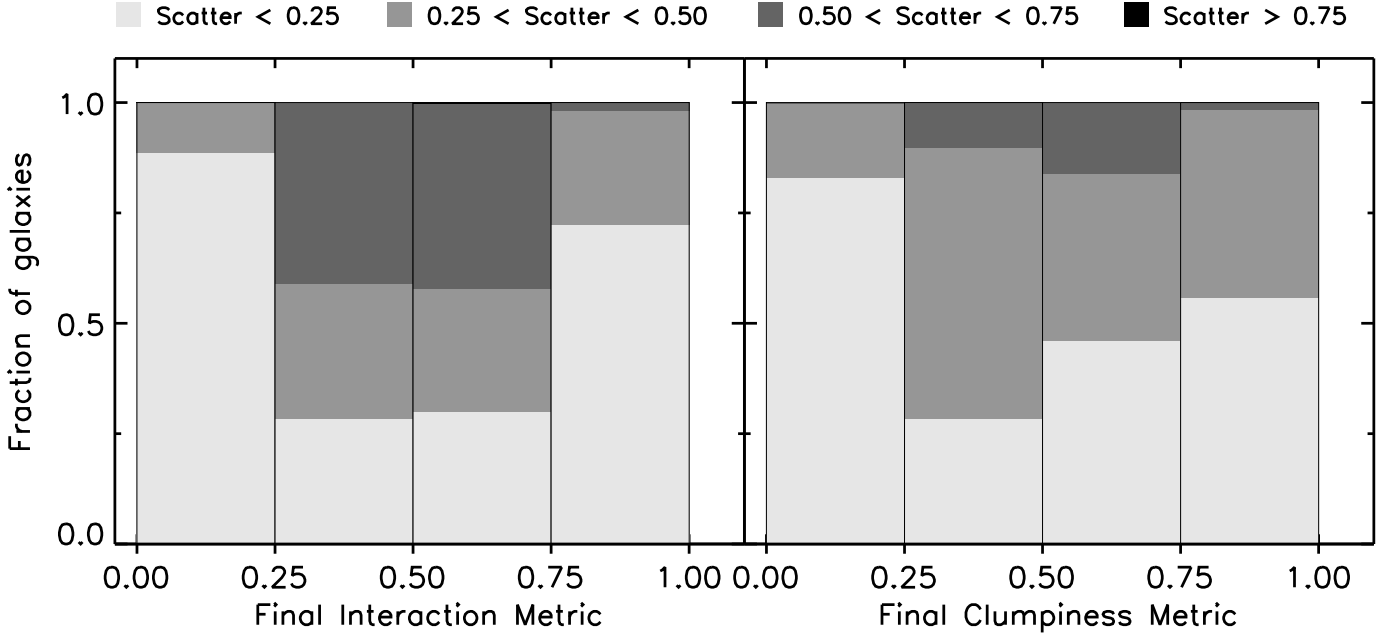}
\caption{The breakup of the mean scatter of visual classification metrics for galaxies at $0.5<z<2.5$ binned by their
final mean metric value. The mean scatter for a galaxy is defined as the average absolute value of the difference of the metric between pairs of classifiers
for that galaxy. Galaxies with larger scatter have a larger level of disagreement between the individual visual classifications.
The relative number of galaxies in four ranges of scatter are shown, represented by lighter to darker grayscale towards increasing scatter.
Galaxies are binned by the final interaction metric (IM, left panel) and the final clumpiness metric (CM, right panel).} 
\label{metric_diffs_breakups}
\end{figure*}

\begin{figure*}[t]
\includegraphics[width=\textwidth]{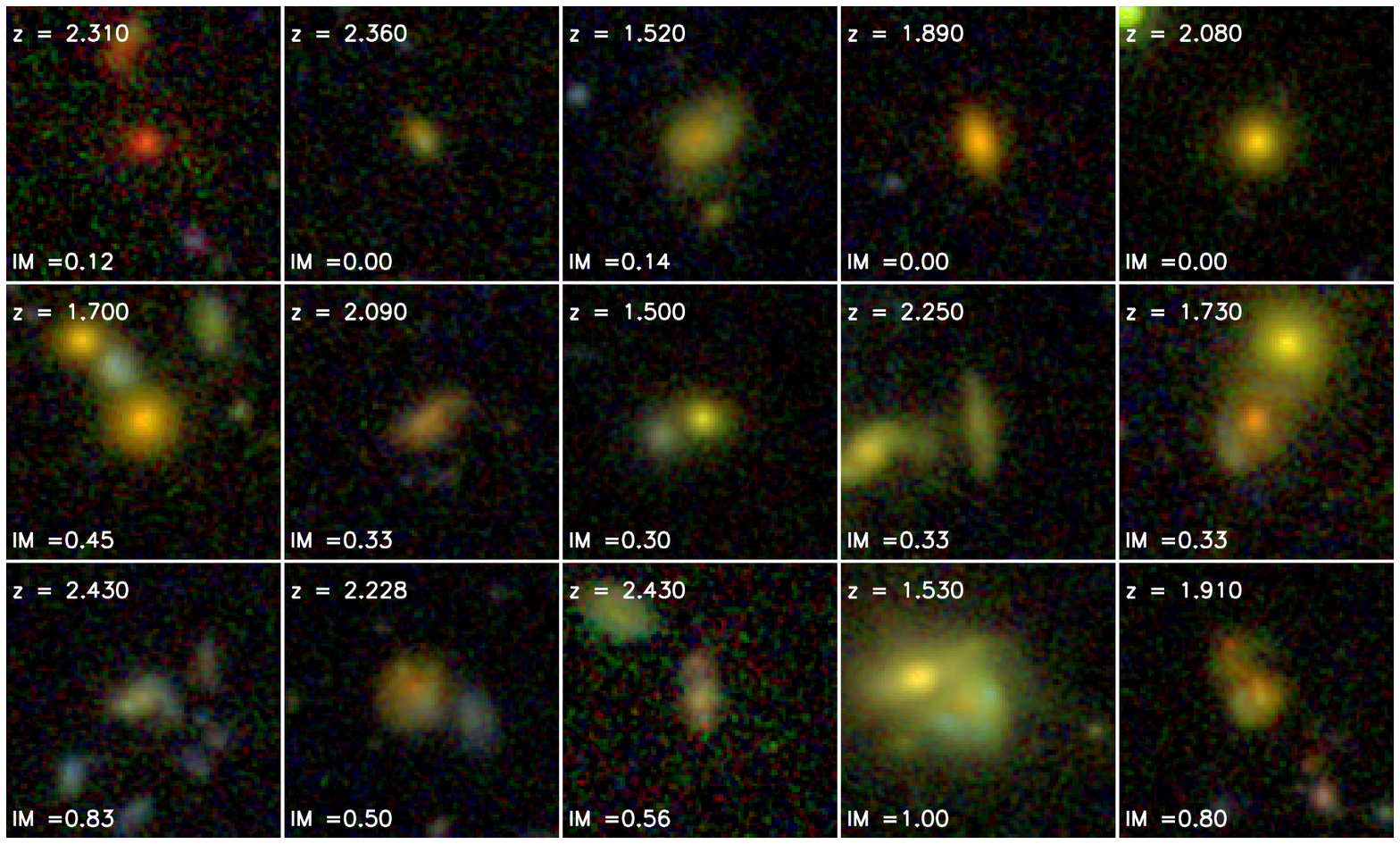}
\caption{A montage of randomly chosen inactive galaxies at $1.5<z<2.5$ arranged in the fiducial bins of the visual Interaction Metric (IM).
Each row of five galaxies are randomly selected from all objects in one of the three bins in IM (see Section \ref{visclass}).
A redshift is written at the upper left of each panel and IM is written at the lower left. All three color (iJH) images are
6" on a side and are identically scaled with a sinh$^{-1}$ stretch.}
\label{im_examples}
\end{figure*}

\begin{figure*}[t]
\includegraphics[width=\textwidth]{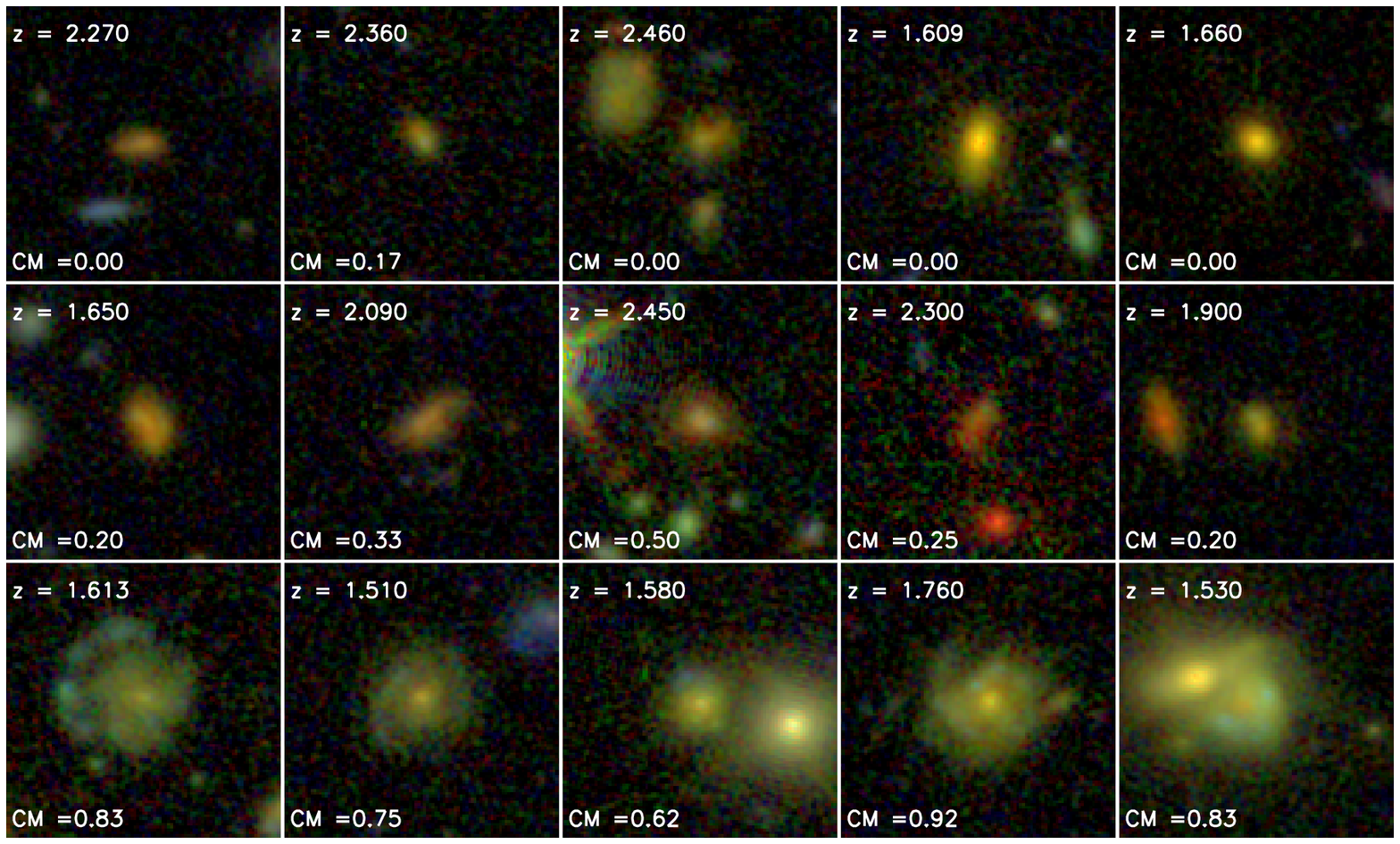}
\caption{A montage of randomly chosen inactive galaxies at $1.5<z<2.5$ arranged in the fiducial bins of the visual Clumpiness Metric (CM).
Each row of five galaxies are randomly selected from all objects in one of the three bins in CM (see Section \ref{visclass}).
A redshift is written at the upper left of each panel and IM is written at the lower left. All three color (iJH) images are
6" on a side and are identically scaled with a sinh$^{-1}$ stretch.}
\label{clmp_examples}
\end{figure*}

\subsubsection{GALFIT models of galaxy structure}

For our primary measure of galaxy structure in the two fields, we rely on light profile fits to the F160W (H-band) images
of galaxies in CANDELS, performed using the GALFIT code \citep{peng10}. Details of the setup and fitting procedure, including
the segmentation of the CANDELS mosaics, object identification, data preparations and initializations, can be found
in \citet{vdwel12}. 

Galaxies are fit using elliptical \sersic\ light profile models, yielding
an estimate of the best-fit major-axis half-light radius \reff, the \sersic\ index ($n$) and the 
ellipse axis ratio ($q$) for each galaxy through GALFIT.
The \sersic\ index is the principal parameter that governs the normalised light profile of galaxy, and, 
in a broad sense, is related to the dynamical state of the stellar matter (though the exact nature of these relationships
and their evolution with redshift is yet unclear). The \sersic\ index is permitted to vary in the range
of 0.5 (sub-exponential) to 8.0 (super-De Vaucouleurs). To ensure the fidelity of the fits, we restrict our study to sources with 
a F160W magnitude $m_{H}<24.0$ and with a GALFIT flag$=0$, indicating a sensible fit.

We also include a set of GALFIT fits to CANDELS F160W and GOODS-S v2.0 F850LP (z-band) images of a subset of 
objects in the CANDELS GOODS-S field, those which overlap with the FIREWORKS multi-wavelength compilation 
\citep{wuyts08}. The GALFIT setup used for these fits is documented in \citet{wuyts11} and is very similar to that used for the H-band fits. 
We compared fits made using both setups for the FIREWORKS subset and found good consistency,
with a median difference in $n$ of 0.04 and a scatter in $n$ of 0.25. About 15\%\ differ by $\Delta n=1$, 
typically towards a higher $n$ using the setup from \citet{vdwel12} and mostly among high \sersic\ galaxies which 
have larger uncertainties in $n$. 

For this work, we fit galaxies, whether active or inactive, using pure galaxy light profile models, without any special components to account
for possible AGN contamination. Studies have shown that excess nuclear light from the AGN in the rest-frame
optical bands at the level of tens of percent can systematically alter single profile galaxy fits, as well as other analytic 
measures of structure. Therefore, earlier studies have tended to include additional point 
source components to AGN profile fits, on the principle that such fits can distinguish between emission from a nuclear point source and the light
of central bulges (which may appear very similar). There has been some calibration of this technique for AGN hosts imaged with
the HST/ACS camera \citep{simmons08, gabor09}. Even with the narrow ACS PSF, such studies found that this method, while generally
sound, overestimates the nuclear point source fraction in a fraction of early-type systems, leading to an enhanced AGN component at the
expense of a residual galaxy component with a lowered \sersic\ index. 
With the broader and more complex HST/WFC3-IR PSF, it is expected that these systematics
would be more pronounced. To avoid being affected by fitting based biases, we employ only single component fits for all galaxies
and use various tests to check for the effects of AGN contamination when presenting results based on the GALFIT fits (Section 3.3.1).

\subsubsection{Visual classification of galaxy structure and disturbance} \label{visclass}

In the CANDELS GOODS-S field, all galaxies with $m_{H}<24.5$ were examined by at least three human classifiers, as part
of a large on-going program of visual classification by the CANDELS team. 
Each classifier looked at four images of the galaxy -- 2-orbit F160W and F125W images, as well as F606W and F850LP images
from the GOODS-S v2.0 release -- along with its SExtractor segmentation map on the F160W mosaic. Several details regarding a galaxy's
visual shape, color, asymmetry, clumpiness and disturbance were noted in a systematic fashion using a GUI tool. 
A set of metrics were used to order the various classifications
down to a final reduced set, taking into account the averaged reliability of classifiers. More details about the plethora of
visual classification outputs, the methodology of classification and a discussion of classifier reliability can be found in \citet{kartaltepe13}.

In this work, we rely on visual estimates of the degree of gross galaxy disturbance, using two different
benchmarks. 
One is the ``interaction metric" ($IM$), a general measure of the level of interaction or merging in a galaxy. 
The CANDELS visual classification tool gives classifiers five choices for each examined object,
along a sequence of increasing apparent degree of interaction. We assign a numerical value to each
choice, which enables us to easily combine the outputs of different classifiers.
Objects which are clearly undisturbed, with no obvious nearby neighbours and no signature of any interaction, are given an $IM=0$. At the other
extreme, objects with $IM=1$ are in obvious late-stage mergers, with highly disturbed structure, strong asymmetries, much clumpiness,
frequently showing tidal tails and/or double or multiple nuclei.  $IM=0.25$ denote objects in an apparent pair or multiple system separated by a few to
several arcseconds with no clear signs of interaction; these may be associated but may also be line-of-sight alignments. $IM=0.5$
are given to objects with visual signatures of interaction with other galaxies outside their H-band segmentation maps -- these may be
pre-mergers or close encounters. $IM=0.75$ are objects that have another galaxy within their segmentation areas and are probably
in an interacting state, but still show distinct structure, likely because the merger is still at an early stage. 
The decisions of individual reliable classifiers may differ, but we average their $IM$ values leading to final $IM$s that could
be intermediate to the five principal assignments described above. The visual IM
for distant galaxies is most sensitive to major mergers, i.e., those where the optical luminosity ratio of the components $\sim1$,
or, alternatively, where the late-stage merger shows strong signs of disturbance.
Minor mergers will only be detectable among the few big bright relatively local galaxies in the CANDELS fields, none of 
which are considered in this work.

To develop a sense of the level of disagreement in the visual assessment of interaction across different classifiers, we consider
the scatter of $IM$ for all galaxies from CANDELS/GOODS-S in the redshift interval $0.5<z<2.5$.
For each galaxy with $N$ classifications, we calculate the mean absolute value of the difference of $IM$ between the $N(N-1)/2$ 
pairs of classifications. Since the number of classifiers for a galaxy could be as low as $N=4$, this quantity is not necessarily a statistically accurate
uncertainty on the final metric, but considering a large number of galaxies together, its distribution allows a simple measure of the variation
in $IM$. We show this distribution in the left panel of Figure \ref{metric_diffs_breakups}, splitting galaxies into 4 bins of their final combined
$IM$ and considering four bins in the mean scatter. It is immediately apparent that galaxies classified as undisturbed or strongly interacting
have only a small scatter among classifiers - most classifiers agree on the two extremes of the visual interaction sequence. Intermediate
values of $IM$ are more uncertain. For example, more than 40\% of galaxies with $0.25<IM<0.5$ have mean
scatters that span across most of the IM sequence ($0.50<$scatter$<0.75$). We are able to trust that a final classification for a galaxy
as non-interacting represents a fair consensus among visual classifiers, but, if a galaxy shows signs of interaction, there may be some uncertainty
as to the degree.

For this reason, we bin galaxies by interaction class quite coarsely: $0.0\leq IM \leq0.2$, $0.2<IM\leq0.5$ and $0.5<IM\leq1.0$, which
represent ``isolated", ``interacting" and ``merging" objects.  Figure \ref{im_examples} shows some examples
of massive galaxies at $1.5<z<2.5$: each row contains randomly chosen galaxies classified into the three coarse
bins listed above.

In addition to $IM$, we identify a refined subset of galaxies with interacting or merging activity, which comes from a combination of visually
identified features derived from the visual classification catalogs. It consists of galaxies 
explicitly listed as visual mergers or direct interactions (disregarding multiple galaxy systems with no clear signs of interactions) 
by at least 60\%\ of classifiers, or in which at least 60\%\ of classifiers noticed tidal tails or double nuclei. 
This multi-feature subset may exclude some very young stage or distant interactions where the galaxies have not yet been 
substantially disturbed, but will be a cleaner distillation of actual systems in a broad array of interaction states, less affected by scatter than the $IM$.  


We also employ a ``clumpiness metric" (CM). CANDELS visual classifiers identified clumpy structure by examining images of galaxies
across multiple bands and were given a choice of three alternatives in the GUI. As for $IM$, we assign numerical values
to these choices. Objects with smooth structure lacking any clumps are assigned $CM=0$. 
Those with one or two clumps (disregarding any central bulge) are assigned $CM=0.5$, 
while objects with three or more clear clumps are given $CM=1$. Intermediate 
values come from averaging the decisions of multiple reliable classifiers. 

In the right panel of Figure \ref{metric_diffs_breakups}, we show the distributions of the scatter in $CM$ calculated in the same
manner as as described above for $IM$. The fidelity of the clumpiness measure of galaxies is generally better than the interaction
measure. Most classifiers agree on whether a galaxy is smooth, but there is a small amount of classifier scatter on the actual level of clumpiness in galaxies.
Figure \ref{clmp_examples} shows some examples
of massive galaxies at $1.5<z<2.5$: each row contains randomly chosen galaxies classified into three coarse
bins in CM ($0.0\leq CM \leq0.2$, $0.2<CM\leq0.6$ and $0.6<IM\leq1.0$).

In addition to IM and CM, we also use flags set by visual classifiers to exclude from our analysis sources with troublesome image
artifacts and, as needed, those with signs of visual point-source contamination. 

It is worth noting that the various measures of structure used here are not independent of each other. Some dependencies
are physical: high \sersic\ galaxies are generally massive, elliptical galaxies which tend to be smooth and, 
even if interacting, frequently show only weak signatures of obvious disturbance such as tidal tails. Some dependencies are
based on the subjectivity of visual classification: if a galaxy is judged to be close enough to another
in the HST images to satisfy a non-zero IM value, a disky/clumpy system frequently appears to be more visually
disturbed and may be assigned a higher IM than a bulgy/smooth system, even if the clumpiness could be related
more to the gas fraction or inflow history of the galaxy rather than its current interacting state. For example, 
one of the high IM galaxies in Figure \ref{im_examples} is also found in the high CM row of Figure \ref{clmp_examples}.
When we compare distributions of individual structural parameters in course of this paper, the reader should keep in mind that there
may be covariances between the various distributions. Disentangling these dependencies is neither trivial nor
fruitful, since the principal limitation in the determination of trends in this work stems from the modest size of the
AGN sample.

\begin{table*}
\caption{Total AGN / Inactive sample sizes after the application of different selections}
\label{selections}
\centering
\begin{tabular}{l c c c c}
\hline \hline
Selection & $0.5<z<1.0$ & $1.0<z<1.5$ & $1.5<z<2.0$ & $2.0<z<2.5$ \\
\hline
CANDELS/both fields + \smass$>10^{10}$ \msun\ (parent) & 91 / 974 &  89 / 877 &  69 / 625  &  34 / 594 \\
parent +  good GALFIT & 87 / 919 &  82 / 841 &  60 / 570  &  22 / 460 \\
\hline
CANDELS/GOODS-S + \smass$>10^{10}$ \msun\ (parent) & 52 / 320 &  46 / 362 &  39 / 211  &  20 / 246 \\
parent +  visual classification & 44 / 293 &  39 / 332 &  33 / 198  &  17 / 229 \\
\hline
\end{tabular}
\end{table*}

\subsection{Herschel/PACS Imaging and Photometry}

Our far-infrared data are composed of maps  at 70 \mics, 100 \mics\  and 160 \mics\ from a 
combination of two large Herschel/PACS programs: the PACS Evolutionary Probe (PEP), a guaranteed time program \citep{lutz11} 
and the GOODS-Herschel key program \citep{elbaz11}. The combined PEP+GH (PEP/GOODS-Herschel) 
reductions are described in detail in \citet{magnelli12}. 
While data at 100 and 160 \mics\ are available in both fields, an additional deep map at 70 \mics\ is also available in GOODS-S.
The PACS 160, 100 and 70 \mics\ fluxes were extracted using sources from archival 
deep Spitzer MIPS 24 \mics\ catalogs as priors, following the method described in \citet{magnelli09}; see also \citet{lutz11}
for more details. 3$\sigma$ depths are 0.90/0.54/1.29 mJy at 70/100/160 \mics\ in the central region of 
GOODS-S and 0.93/2.04 mJy at 100/160 \mics\ in GOODS-N. The GOODS-S maps are $\approx 80$\%\ deeper 
than the GOODS-N maps and probe further down the FIR luminosity function at all redshifts \citep{magnelli12}.

For practical purposes, we use the monochromatic luminosity of a galaxy at 60 \mics\ rest (\lfir) 
as a measure of its FIR luminosity. 
The PACS bands cover the rest-frame 60 \mics\ over much of the redshift
range probed in this work and we estimate \lfir\ from a simple log-linear interpolation of PACS measurements
in bands that bracket 60 \mics\ in the rest-frame. The use of \lfir\ obviates the need to apply an uncertain
correction between monochromatic and total FIR luminosities. As a rough guide for the reader,
a star-forming galaxy with an IR SED similar to M82 with \lfir$=10^{45}$ \lsun\ has a SFR of approximately
80 \msun/yr. The exact transformation depends on the SED shape, the SF history of the galaxy and
on many other factors relating to the distribution of dust and stars in a system.

In cases where a mean \lfir\ is desired for a sample
consisting of a mix of PACS detected and undetected sources, we follow a technique developed in earlier works
from our team \citep{shao10, santini12} and briefly outlined here.
We bin sources in this study by redshift and structure. We employ fiducial redshift bins: $0.5<z<1.0$, $1.0<z<1.5$,
$1.5<z<2.0$ and $2.0<z<2.5$ (though, in some analyses, we combine the last two redshift bins to improve statistics).
A fraction of sources in each bin are detected in two or more PACS bands. \lfir\ is
calculated for these using their individual redshifts and a log-linear interpolation of PACS fluxes. Of the remaining sources,
some are detected in only one PACS band, while the majority are undetected in the FIR data. For the latter, stacks were
performed at the optical positions of the sources on PACS residual maps in all available bands,
from which mean fluxes were measured using PSF photometry. The stacked fluxes in a band were averaged with the fluxes of
sources only detected in that band, weighting by the number of sources in each category. This gives mean fluxes for the 
partially detected and undetected AGNs in both bands, from which a mean \lfir\ was derived using the median redshift of 
these sources. The final 60 \mics\ luminosity in each bin was computed by averaging over the linear luminosities of 
detections and non-detections, weighted by the number of sources. This procedure was only performed for bins 
with more than 3 sources in total.

Errors on the infrared luminosity are obtained by bootstrapping. A set of sources
equal to the number of sources per bin is randomly chosen 100 times among detections and non-detections (allowing repetitions), and
 \lfir\ is computed per each iteration. The standard deviation of the obtained \lfir\ values gives the error on the average 
 60 \mics\ luminosity in each bin. The error bars thus account for both measurement errors and the scatter in the population distribution.

\subsection{Chandra Deep Field (CDF) X-ray catalogs} 

Cospatial with the two GOODS survey fields \citep{giavalisco04}, the Chandra Deep Fields (CDFs) are the deepest pencil-beam X-ray surveys in the sky. The 2 Msec exposure in the full CDF-North (CDF-N) has produced a point source catalog 
consisting of 503 sources \citep{alexander03}, while in GOODS-South, the recent
4 Msec CDF-South (CDF-S) catalog comprises 740 sources \citep{xue11}. However, only $\sim 60$\% of the full CDFs are imaged
by CANDELS.

We have extensively characterized the data and 
catalogs in both fields, in which careful associations have been made with optical and near-IR counterparts,
using, where possible, probabilistic crossmatching models \citep{luo10, xue11}. In addition to 
the deep X-ray data, the wealth of deep spectroscopy and multi-wavelength photometric data in the GOODS fields
have enabled accurate spectroscopic or AGN-optimised photometric
redshifts to be determined for the majority of the X-ray sources \citep[e.g.,][]{szokoly04, luo10}. We estimate absorption-corrected
hard-band X-ray luminosities ($L_X$) of sources with redshifts using spectral modeling techniques \citep{bauer04}.
As a result of the small area and great depth of the CDF exposures, most X-ray sources are low or moderate
luminosity AGNs -- only $\sim 5$\%\ of the sources have $\log L_{X} \textrm{(2-10 keV)} > 44$ \ergs. These may be
thought of as X-ray selected equivalents of the local Seyfert galaxy population.
In this work, we only consider sources with $\log L_{X} > 42$ \ergs, to prevent contamination from powerful starbursts, in which
emission from high-mass X-ray binaries can potentially overpower the emission from nuclear activity.

\begin{figure*}[t]
\includegraphics[width=\textwidth]{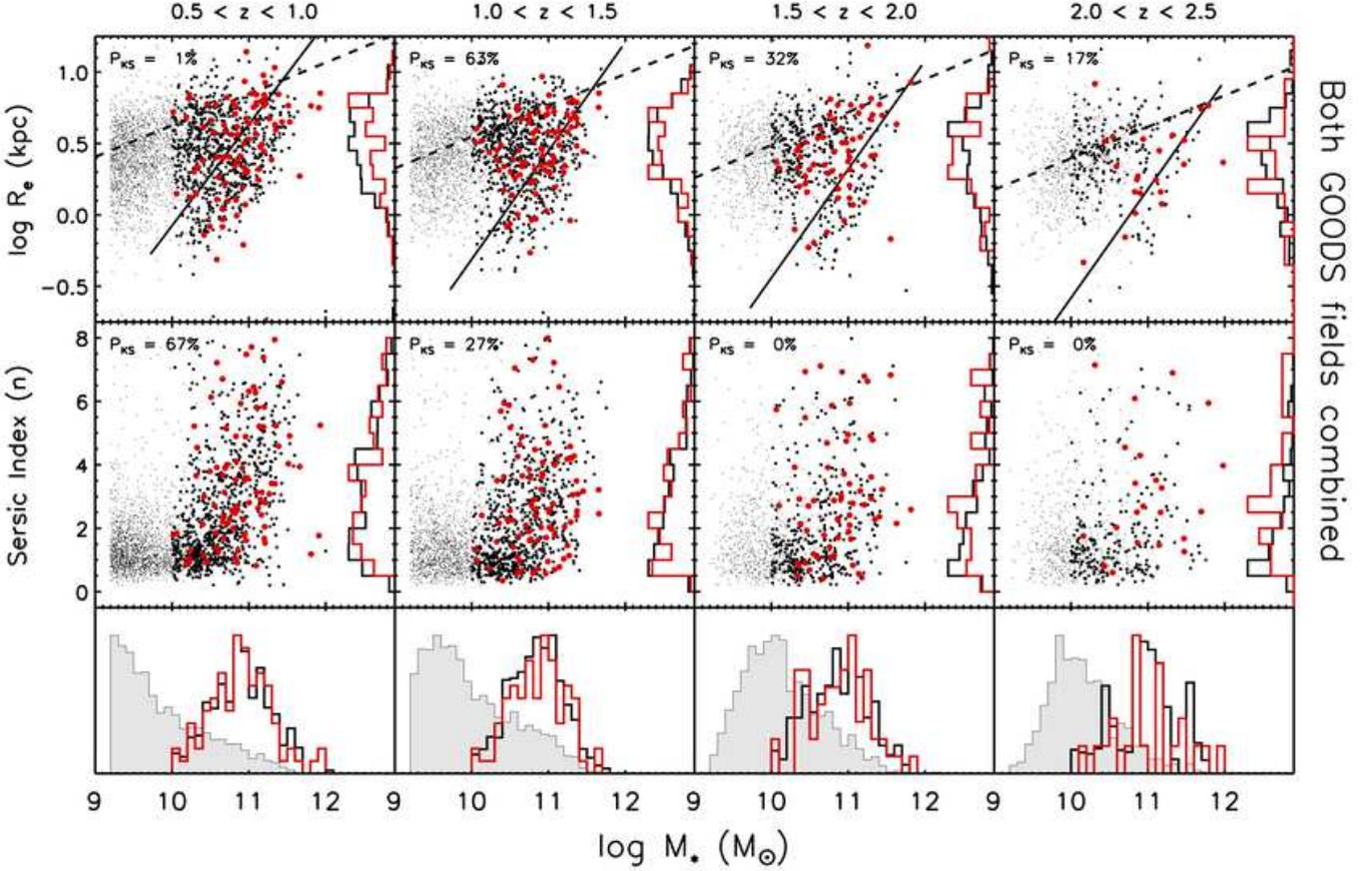}
\caption{{\bf Top panels:} Physical half-light semi-major axis radii plotted against stellar mass \smass\ for AGNs (large red points),
inactive control galaxies (large black points) and the overall galaxy population (small grey points), in four distinct redshift bins. 
Samples from both GOODS fields are combined and one hundred control sample draws are used in all panels of this Figure.
The lines in the upper panels show parameterisations of galaxy size-mass relationships across redshift from \citet{vdwel14};
dashed lines for late-type galaxies and solid lines for early type galaxies.
Vertical histograms compare the size distributions of the AGNs (red lines) and the inactive control.
{\bf Middle panels:} \sersic\ index $n$ plotted against stellar mass \smass\ for AGNs (large red points),
inactive control galaxies (large black points) and the overall galaxy population (small grey points), in four distinct redshift bins. 
Samples from both GOODS fields are combined here. Vertical histograms compare the size
distributions of the AGNs (red lines) and the inactive control. In each panel in both rows, the Kolmogorov-Smirnoff (KS) 
probability ($P_{KS}$) that the vertical distributions come from the same parent sample is shown in the upper left corner.
{\bf Bottom panels:} A comparison of the \smass\ distributions of AGNs (red histograms), the inactive control sample
(black histograms) and the overall galaxy population (grey filled histograms). The importance of stellar mass-matching
in constructing a valid comparison sample is clear, and the performance of our mass-matching procedure
may be appreciated at a glance. 
}
\label{galfit_trends}
\end{figure*}

\subsection{Multi-wavelength photometry and stellar masses}

We employ multiwavelength galaxy catalogs in both GOODS fields to define a general galaxy sample, the
properties of which we will compare to the AGNs. In GOODS-S, we use the updated 
GOODS-MUSIC database \citep{santini09, grazian06}, while in GOODS-N we use a catalog 
developed for the PEP team using a similar methodology \citep{berta10, berta11}
The former catalog selects galaxies 
with observed magnitudes in the HST F850LP band $<26$ or in the ISAAC $K_s$ band $<23.5$, while the 
latter is primarily selected to have $K<24.2$. In order to exclude a surfeit of faint sources with inaccurately red
colors and masses, we apply an additional cut of F850LP $<26$ in the GOODS-N catalog. For galaxies with no current
spectroscopic redshifts, photometric redshifts were determined by fitting multiwavelength photometry using
PEGASE 2.0 templates \citep{fioc97} in GOODS-S or using the EAZY code \citep{brammer08} 
in GOODS-N. For details on the catalog preparation, characterization and photometric redshift estimation, we
refer the reader to \cite{santini09} for GOODS-S and \cite{berta10} for GOODS-N.

We have developed a custom technique to estimate the stellar masses (\smass) in AGNs, 
by linearly combining galaxy population synthesis model templates and AGN SED templates to fit multiwavelength
photometry. For inactive galaxies, we perform a
$\chi^{2}$ minimization of \cite{bc03} synthetic models, assuming a Salpeter Initial Mass Function (IMF) and parameterizing
the star formation histories as exponentially declining functions. For AGNs, we also include
an AGN template from \cite{silva04}, which accounts for a variable fraction of the total light of the galaxy. The AGN template
reflects the classification of the X-ray source, derived from information about its SED and
spectrum, where available. For sources classified as Type I (broad lines in the spectrum, clear AGN contribution
in the rest-frame optical and UV), an unobscured AGN SED was used. For the rest, a Type II template was used
if the estimated X-ray absorption column $N_H < 10^{24}$ cm$^{-2}$, and a Compton-thick template for
more heavily absorbed systems, though in practice, the choice of the latter two templates makes little difference. 
For further details of the method, performance evaluations, tests and limitations,
we refer the reader to \cite{santini12}. 

While AGN are selected by their X-ray emission, we define our `inactive' galaxy population as all galaxies that are
undetected in X-rays (excluding even those which have  $\log L_{X} < 42$ \ergs) and 
Spitzer/IRAC colors that are unlike those of bright AGNs following the criteria of \cite{donley12}. In practice, only a very small
fraction of the general galaxy population are rejected on the basis of these criteria, but they tend
to be massive galaxies and could potentially sway the statistics of SF comparisons among such systems 
by an inordinate degree. We also impose a minimum mass of \smass$=10^{10}$ \msun\ on both AGNs and inactive
galaxies. Very few AGNs in our redshift range of interest lie at lower masses and the GOODS galaxy catalogs
become increasingly incomplete below this mass limit at $z\sim2$ \citep{santini12}.

Galaxies from the multiwavelength catalogs were matched to Herschel sources through the positions of the 24 \mics\
priors, which are, in turn, tied to IRAC catalogs in the GOODS fields \citep{magnelli09}. 
Therefore, the cross-matching tolerances are $\approx 1"$. The rate of spurious crossmatches is $\approx3$\%, estimated from the asymptotic
behaviour of the cross-match offsets between the IRAC catalog and the CANDELS H-band catalog in GOODS-S.

In Table \ref{selections}, we list the numbers of AGNs and inactive galaxies from the parent samples -- sources that lie
in the overlap of the CANDELS imaging and Herschel/PACS maps, additionally restricted by the footprint of the CDFs 
(for AGNs) or the multiwavelength catalogs (for the inactive galaxies). Along with the parent sample numbers, we also list numbers of AGNs and inactive
galaxies that satisfy our criteria for good GALFIT measurements (in both GOODS fields) and good visual structure estimates (in GOODS-S only).

\begin{figure*}[t]
\includegraphics[width=\textwidth]{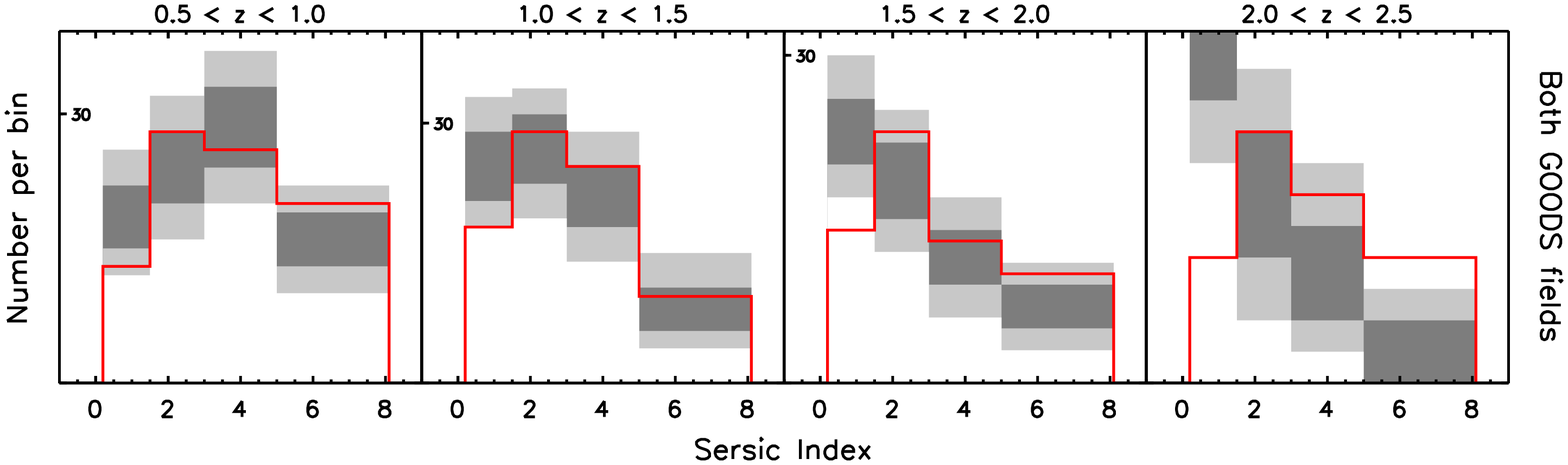}
\includegraphics[width=\textwidth]{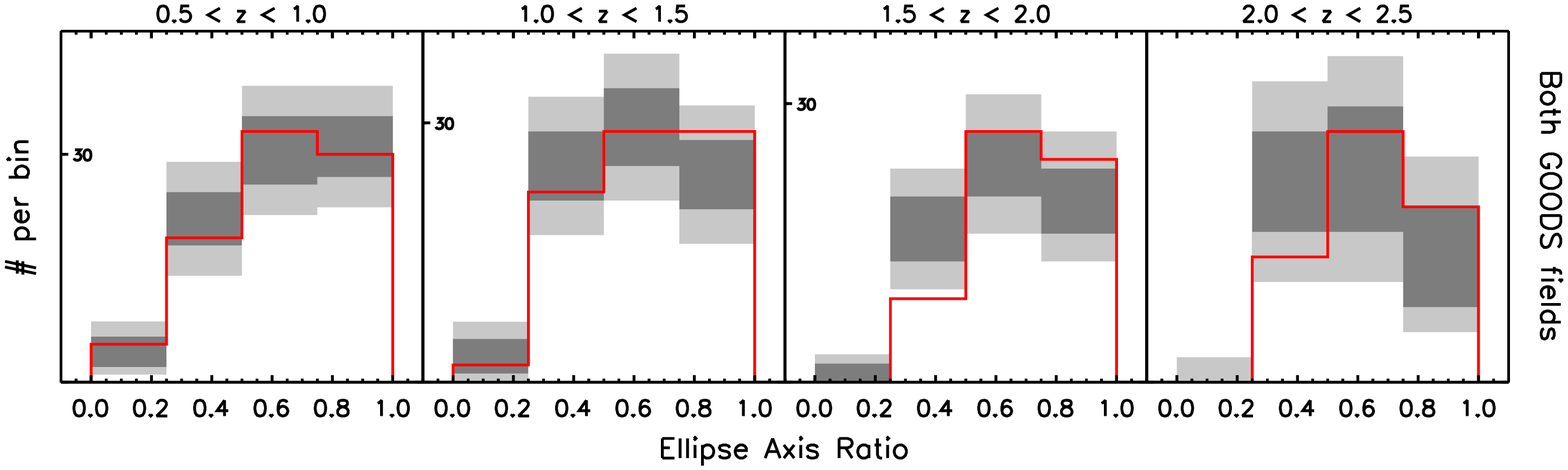}
\includegraphics[width=\textwidth]{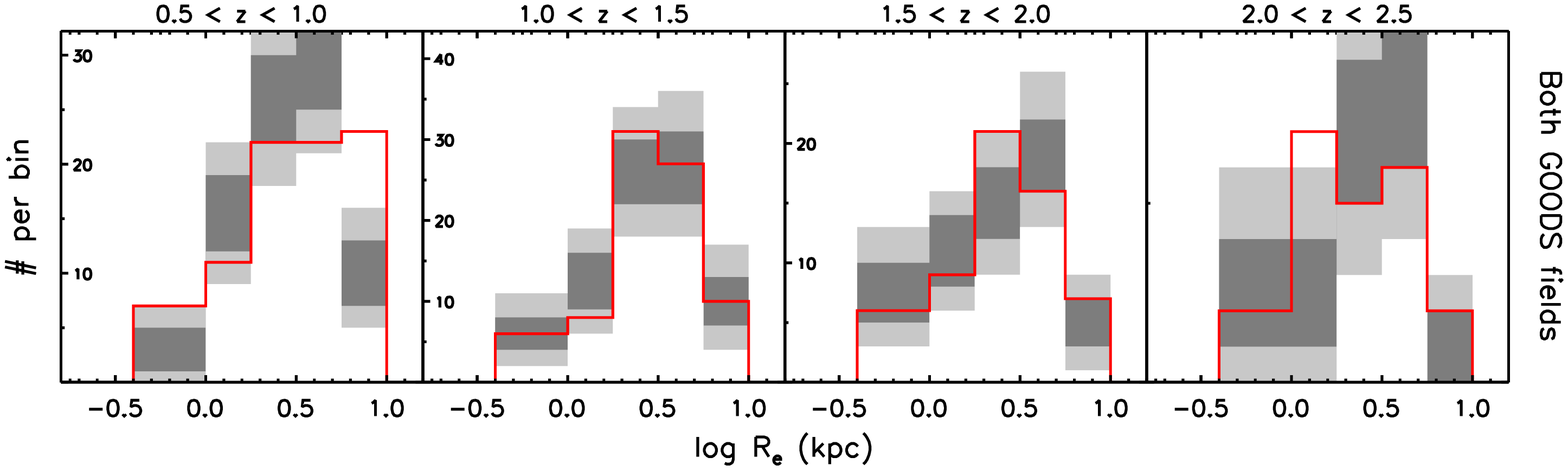}
\caption{Distributions of the light profile parameters from GALFIT fits for AGNs and inactive control galaxies. The
three rows of panels show histograms in \sersic\ index ($n$), ellipse axis ratio ($q$) and 
physical semi-major axis radius (\reff) from top to bottom respectively. Panels left to right span four distinct
bins in redshift. Samples from both GOODS fields are combined. 1000 draws of a mass-matched control sample are analysed to determine 
the $1\sigma$/$2\sigma$ scatter in the distributions of the parameters for inactive galaxies, shown as dark/light
grey zones in the histograms. Red open histograms show the parameter distributions for the AGNs.}
\label{galfit_distributions}
\end{figure*}

\subsubsection{Bootstrapped stellar mass-matched control samples}

Moderately luminous AGNs are inherently rare among galaxies. In the small fields considered in our study, only few to several
tens of AGNs that satisfy our the minimum criteria for a valid structural measurement may be found in each redshift bin. 
Direct constraints on the relationships between SFR, redshift and structure based solely on the AGNs themselves are severely limited by small number
statistics. 

On the other hand, the inactive galaxy population far outnumber X-ray selected AGNs in both fields and at all redshifts. We
take advantage of this by creating multiple control samples of inactive galaxies matched to the AGNs using Monte-Carlo bootstrap
techniques, from which we constrain the measurements and distributions of structural and SF properties more rigorously, as well as account for 
the uncertainties arising from the small sample size of the AGNs and complex scatter associated with the structural measurements. The 
observed measurements and distributions shown by the AGNs may then be compared to those of the control samples. 
Since the uncertainties in the latter are also estimated,
we can ascertain whether or not the observed AGNs are consistent with being drawn from the inactive galaxy population. 

We match galaxies to AGNs on the basis of stellar mass. All AGNs and inactive galaxies in each of the fiducial redshift bins are
further binned into narrow mass intervals, of $\Delta$\smass$=0.2$ dex for \smass$<10^{11.5}$ \msun\ and 
$\Delta$\smass$=0.5$ dex for \smass$>10^{11.5}$ \msun. The increase of the matching tolerance at high \smass\
is due to the paucity of high mass galaxies, as well as the high
AGN incidence at these masses, which reduces the pool of control galaxies substantially.
For each AGN in a redshift and mass bin, we randomly choose one counterpart from the corresponding
set of inactive galaxies, allowing for duplications. This yields a single set of inactive control galaxies of the same
number as the AGNs, sharing their stellar mass distribution within the matching tolerance. We repeat this process hundreds of times
to get multiple independent control samples. These are used to determine a ``statistical distribution'' of any parameter of interest, which
encapsulates both the mean distribution and the statistical scatter in the distribution coming from real and sampling
variance of both the inactive galaxy population and the measurements of the parameter itself. We then evaluate whether
the distribution of the parameter for the AGNs is consistent with arising as a single draw of the control sample. In this sense,
the approach taken in this work is purely comparative and no attempt is made to correct the distributions for incompleteness
or biases associated with the photometric cuts and the quality cuts applied for structural measurements. Since both AGNs 
and galaxies are tied to the same multi-wavelength catalogs, the same biases are expected to apply to both
populations.

Throughout the rest of this paper, we will use the term ``inactive galaxies" to specifically denote the population of massive
galaxies without detectable nuclear activity with a stellar mass distribution matching those of the AGNs, i.e., the stellar mass-matched
control samples described above.

\section{Setting the stage: Structural patterns for AGNs and inactive galaxies}

We start by highlighting a few important structural relationships inherent for galaxies, as uncovered by our morphological
measurements. These serve as context for understanding differences between AGNs and inactive galaxies
in terms of their relationships between SF and galaxy structure. 

\begin{figure*}[t]
\includegraphics[width=\textwidth]{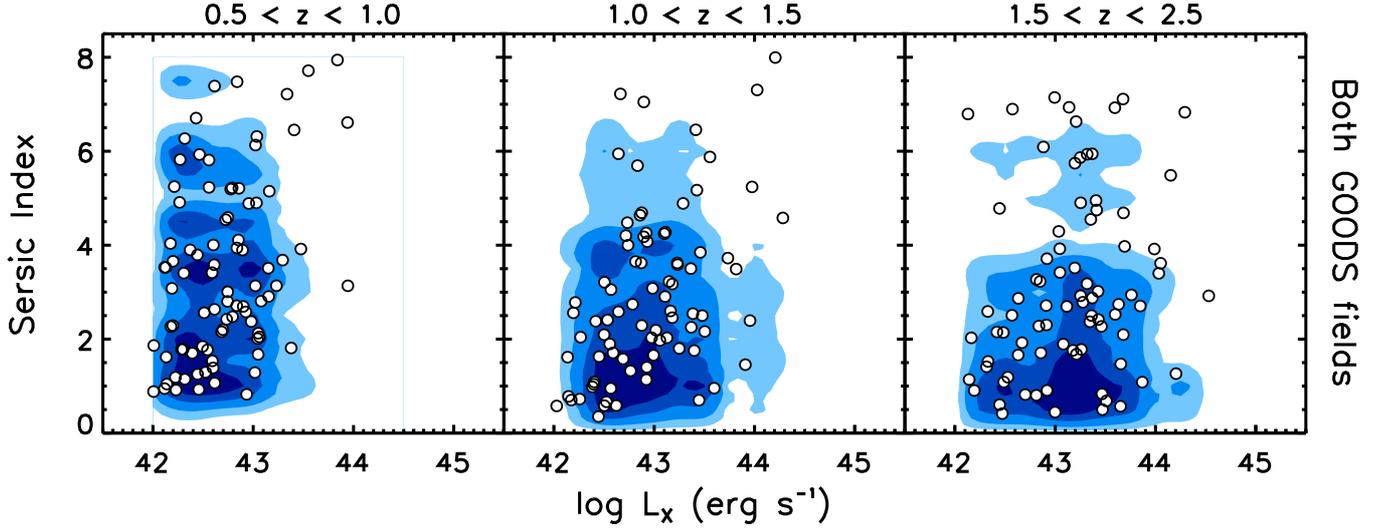}
\caption{GALFIT-derived \sersic\ index vs. absorption-corrected 2-10 keV X-ray luminosity of AGNs from both Chandra Deep Fields
combined (white circle points). The contours show the density of points from 1000 draws of a mass-matched control sample of inactive
galaxies. To place the control galaxies on this plot, the X-ray luminosity assigned to each control galaxy 
is that of the AGN to which it is matched, producing, in effect, a large ``simulated AGN" sample with 
the same mass and X-ray luminosity distributions as the real AGNs. The four contour levels
encompass 90\%, 75\%, 50\% and 25\% of the simulated AGNs in this plane.}
\label{lx_sersic}
\end{figure*}

\subsection{Size--mass relationship}

Galaxies show a clear trend towards larger sizes at higher stellar mass, a relationship that is discernible to $z>2$. 
We consider the size-mass relationship in the upper panels of Figure \ref{galfit_trends}, 
where we plot H-band half-light radii \reff\ vs.~\smass\ for galaxies and AGNs combining the two GOODS fields. 
As a guide to the eye, lines are used to illustrate the typical size-mass relationships 
for star-forming galaxies and quiescent galaxies as recently determined from the CANDELS-based study of \citet{vdwel14}.
Since current star-formation is correlated with the light/mass profile of galaxies, early- and late-type galaxy populations with 
different \sersic\ indices ($n$) are also differentiated by their characteristic size-mass relationships. 
High $n$ galaxies show a much steeper relationship between size and mass and are generally smaller 
than low $n$ galaxies at the same \smass. In addition, the size-mass relationship of high $n$ galaxies evolves
more strongly to $z=2$ \citep{trujillo06, trujillo07,bruce12}. The differences between these relationships and their 
evolution place important constraints on the buildup of stellar mass in galaxies from $z=2$. 

The AGNs and 100 sets of mass-matched control galaxies are shown in this diagram as larger red and black points respectively.
For the most part, AGNs occupy the same range and scatter in size as shown by inactive galaxies of the same stellar mass. This is shown
more clearly through the vertical histograms on the side of each panel, which compare the size distributions
of the AGNs and the control galaxies. The similarity of the distributions is measured using a two-sided
Kolmogorov-Smirnoff (KS) test and the probability that the two distributions are drawn from the same parent population ($P_{KS}$)
is listed in the upper left corner of each panel. We see some systematic variation in the relative size distributions with redshift.
The KS tests suggest that the two distributions are very consistent between 
$z=1$ to $z=2.5$, but differs in the lowest. At $0.5<z<1.0$, AGN hosts are larger than the control sample.
However, a detailed look in Section 3.3 suggests only minor differences, which may be driven by cosmic variance.

\subsection{\sersic\ index--mass relationship}

The presence of two size-mass relationships among galaxies is related to differences in their light profiles, as parameterised by
the \sersic\ index ($n$) \citep{trujillo06} This is believed to be driven by their different dynamical evolution \citep[e.g.][]{baugh96,mo98,naab09}. 
Galaxies with $n\approx1$ have primarily exponential light profiles and are likely to be disk-dominated. Since disks are dynamically
colder systems with substantial rotation, low \sersic\ galaxies are unlikely to have suffered a violent event 
within several dynamical times of our view of them. On the other hand, galaxies with $n\approx4$ are classical spheroids, which
are dynamically hot and have low rotation. This suggests that their most critical
recent evolutionary event was violent enough to redistribute any cold components that may have existed in the progenitors
of these galaxies. This, and the relative importance of minor mergers and recent gas accretion, play a role in the interpretation
of the different size-mass relationships shown by low and high \sersic\ galaxies. Some complexity to this picture comes from
suggestions that modest \sersic\ indices could be achieved by disk-dominated galaxies in the high redshift
Universe \citep[e.g.][]{vdwel11}.

In the middle panels of Figure \ref{galfit_trends}, we plot $n$ vs.~\smass\ for galaxies and AGNs for the two
GOODS fields combined. At \smass$<10^{10}$ \msun\ (with possibly some differences over redshift), the vast majority of galaxies
have low \sersic\ indices of $n<2.5$. Towards \smass$>10^{10}$ \msun, the typical \sersic\ index of galaxies increases considerably,
reflecting the increased spheroidal dominance of high mass galaxies seen both in the local and distant Universe \citep[e.g.,][]{buitrago13}.
There is clear change in the fraction of high $n$ galaxies with redshift over the range we study here. At $0.5<z<1.0$
and \smass$>10^{10}$ \msun, 35\% of galaxies have $n>2.5$, which changes to 18\% among equally massive galaxies
at $2.0<z<2.5$. This is a well-established result in the CANDELS era \citep[e.g.][]{vdwel11}.

As before, we show the AGNs and control galaxies using larger red and black points, and compare their \sersic\ index
distributions visually using vertical histograms in each panel and analytically with two-sided KS tests. At $z<1.5$, 
AGNs and inactive galaxies have statistically indistinguishable distributions. However, at higher redshifts, AGNs
show a significant surfeit of high \sersic\ systems. In the next subsection, we examine the differences between
AGNs and inactive galaxies more carefully to gain some insight into the causes for these observed differences.

\begin{figure*}[t]
\includegraphics[width=\textwidth]{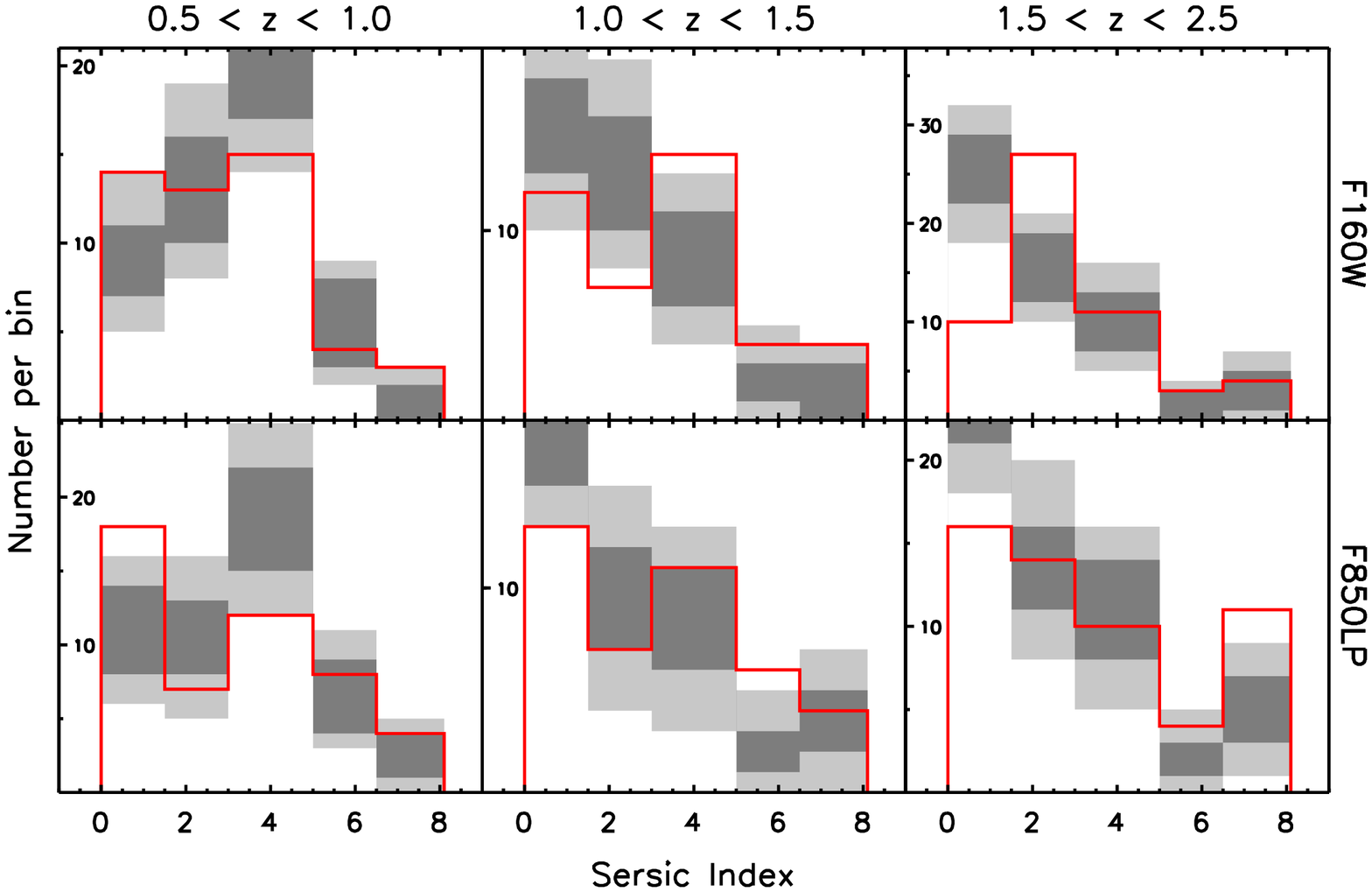}
\caption{Distributions of the \sersic\ index ($n$) from GALFIT fits for AGNs and inactive control galaxies in two HST
bands: the ACS/F850LP band (bottom panels) and the WFC3/F160W band (top panels).
Panels left to right span three distinct bins in redshift. Only the subset of sources contained in the FIREWORKS
catalog are here. 1000 draws of a mass-matched control sample are analysed to determine 
the $1\sigma$/$2\sigma$ scatter in the $n$ distributions for inactive galaxies, shown as dark/light
grey zones in the histograms. Red open histograms show the parameter distributions for the AGNs.
\sersic\ indices are typically lower in the bluer F850LP band. Especially at $1.5<z<2.5$, the $n$ histograms of the
AGNs are more consistent with those of the inactive galaxies from the F850LP fits.}
\label{sersic_comps}
\end{figure*}

\begin{figure}[t]
\includegraphics[width=\columnwidth]{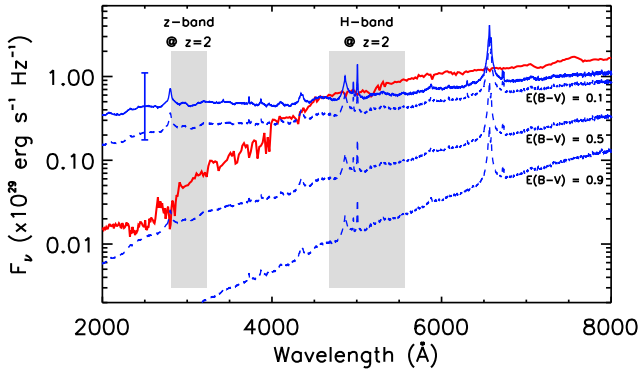}
\caption{A comparison of a typical galaxy spectrum (red solid line) and AGN spectra with different levels of extinction (blue lines).
The base galaxy spectrum is  the `Sa' template from the SWIRE library \citep{polletta07} scaled to correspond to $H=24$ magnitude if
placed at $z=2$.
The base AGN spectrum (blue solid line) is stitched together from composite radio-quiet QSO spectra in the UV from \citet{telfer02}
and the optical from \citet{vandenberk01}, scaled to correspond to \lhard$=10^{43.5}$ \ergs\ using optical/X-ray relation
defined by \cite{lusso10}; the error bars show the expected scatter in the relation. 
The blue dashed lines show the effects of Calzetti-law extinction on the base AGN spectrum
with E(B-V) as indicated on the right below the corresponding extinguished spectrum. The approximate position of the observed
F850LP (z) and F160W (H) bands at $z=2$, relevant for the interpretation of Figure \ref{lx_sersic}, are shown as grey bands.  
}
\label{spectral_checks}
\end{figure}

\subsection{Light profile parameter distributions}

The large number of inactive galaxies from the parent catalogs enables the construction of statistical
distributions of $n$, \reff\ and the ellipse axis ratio $q$ for the mass-matched inactive control sample, combining
both GOODS fields to maximise the sample size and minimize cosmic variance. 
In Figure \ref{galfit_distributions}, each row shows histogram distributions for a different light profile
parameter output by GALFIT. In each panel, the dark/light grey regions show, for each histogram bin, 
the 1$\sigma$/2$\sigma$ range in the number of inactive galaxies in that bin, determined from 1000 control sample draws. 
The distribution for the AGNs are shown as red open histograms. If the AGN histogram lies within the dark grey regions, 
their distributions are completely consistent with those of inactive galaxies within 1 standard deviation.

By and large, the distributions of all the structural parameters among AGNs are not radically dissimilar from the
control sample across all redshifts. 
Closer scrutiny suggests some systematic differences. The distributions of \sersic\ index of the AGNs, 
are broadly consistent with those of the control galaxies at $z<1.5$ but deviate significantly towards higher $n$
at $z>1.5$. At these high redshifts, the shapes of the AGNs are a bit rounder, with a $q$ distribution skewed slightly 
towards higher values than the inactive galaxies. AGNs also tend to be a bit larger than inactive galaxies
at $0.5<z<1.0$ and a bit smaller than inactive galaxies at $1.5<z<2.5$. 

An interesting trend to note is that the light profile distributions of AGNs change very little across redshift, unlike that of the inactive
galaxies. Between $z=0.5$ and $z=2.5$, the \sersic\ index distribution of the AGNs has a roughly fixed median value of 
$\approx2.7$, while that of the inactive galaxies ranges from $\approx3.0$ at $0.5<z<1.0$ to $\approx1.5$ at $2.0<z<2.5$.
In parallel, inactive galaxies also become more elongated, a sign of the increased prominence of massive disks with redshift. AGNs, however,
remain rounder, with only a slight shift towards more elongated profiles between the low and high redshift bin. The median sizes
of the control galaxies increases slightly with redshift, while those of the AGNs decreases slightly with redshift

Taken together, these systematics can be attributed to two possible effects. 
A small amount of point source contamination from the nucleus may be affecting AGN light profiles, mostly
in the high redshift bin where the AGNs are more luminous and the H-band traces the rest-frame B and V bands ($4400$--$6000$ \AA).
Alternatively, there may be a preference for AGNs to be found in more bulge-dominated hosts, as
suggested by earlier visual classification studies \citep{kocevski12}. Both effects lead to a greater concentration of light
coming from the central region of the galaxy, resulting in the small differences in light profile structure we observe.

\subsubsection{Possible origins of the central light excess}

An AGN can be a profuse source of energy emitting across most wavelengths. Unextinguished UV-to-optical 
continuum emission from a nucleus originates from the hot accretion disk and is very blue, rising
in flux rapidly at shorter wavelengths towards a putative peak in the extreme UV. On this
continuum may be superimposed emission lines from the various AGN-ionized regions. Regarding
its effect on the appearance of galaxy structure, unextinguished AGN appear as nuclear point sources
with blue colors. However, dust along the line of sight to the nucleus can redden
the point source, making it harder to distinguish against the light from the bulge of the host galaxy.

The influence of strong AGN point sources on galaxy structural measurements can be quite profound.
Even point source fractions as low as a few tens of percent can lead to appreciably
bulgier, rounder and more centrally concentrated effective light profiles \citep{pierce10}. 
It is worth noting, given the nuclear luminosities of the AGNs in the GOODS fields (as estimated
from their X-ray emission), only a small fraction should be powerful enough to severely contaminate
their host galaxy light, even if unextinguished by dust (see the Appendix of \citet{rosario13a} for a discussion).
Earlier GALFIT studies have revealed frequent red nuclear excesses among AGNs in the GOODS field
\citep{simmons11,schawinski11}, attributable to obscured AGN point sources. These sources
are generally much fainter than their hosts and would not strongly alter their measured structure. However,
pervasive weak nuclear excesses among AGNs could explain some of the mild systematic trends we find in the
light profile parameter distributions of AGN hosts.


In Appendix A, we undertake a brief study of the evidence for nuclear contamination of the light
profiles of AGNs at $1.5<z<2.5$. We fit AGNs and mass-matched inactive galaxies using a combination
of a \sersic\ galaxy model and a central PSF. The inclusion of a point source in the GALFIT fits leads to
a lower effective \sersic\ index distribution for the AGN hosts (Figure \ref{psf_test1}), but it also systematically
lowers the effective indices for inactive galaxies. As strong nuclear emission is not expected in the latter population,
we conclude that such two component fits are subtracting away the light of a central bulge as well. This is not
unreasonable given the similar sizes of bulges and WFC3/F160W PSFs at $z\sim2$ \citep[e.g.][]{bruce12}.
An accurate treatment of the influence of AGN nuclear emission requires multi-component fits, as well as
careful simulations with the addition of fake point sources to real images of massive inactive galaxies 
in order to understand the systematics inherent in the consideration of a second structural component in GALFIT.
This will be pursued in future work built upon this study. We can, nonetheless, perform some simple
tests for the importance of strong blue excesses to the light profiles of AGNs.

In Figure \ref{lx_sersic}, we plot the \sersic\ indices of AGNs against their absorption-corrected 2-10 keV
X-ray luminosity \lhard, combining the samples from both GOODS fields. 
The underlying contours show the distribution in this plane of 1000 draws of a mass-matched
control sample of inactive galaxies. To be placed in this Figure, each control galaxy is assigned the X-ray luminosity of the 
AGN to which it is matched, generating, in effect, a large ``simulated AGN" sample. 
If the light profiles of AGNs were independent of nuclear luminosity and identical to those of inactive galaxies, 
they are expected to scatter in this plane in the same fashion as the control sample. This is the case at $z<1$.
At higher redshifts, there is a definite tendency for the more luminous AGNs to occur in systems with higher $n$.
In particular, one may note that AGN hosts with $n>4$, which may be seen at $z<1$ across \lhard, are more common in sources with
 \lhard$\gtrsim  10^{43}$ \ergs\ in the two higher redshift samples. This Figure seems to suggest that the more
 luminous AGN hosts at $z>1.5$ are associated with high \sersic\ systems, perhaps due to higher levels of AGN contamination,
 or because such luminous AGNs are preferentially found in hosts with larger spheroids. A similar result was found in 
 the visual study of X-ray AGN hosts in CANDELS/GOODS-S by \cite{kocevski12}.
 
We perform another simple test of these alternatives by employing GALFIT fits of the same galaxies in a bluer band. If blue
nuclear contamination is the reason for the systematic deviation of $n$ in AGN hosts at $z>2$, and assuming
that AGN spectra do not change systematically with redshift, then we should see a qualitatively similar deviation
at $0.5<z<1.0$ when we examine $n$ distributions from F850LP (z-band) fits, 
since this band traces the rest-frame B-band for these galaxies. 
In addition, the deviation should be much more pronounced for the higher redshift systems where F850LP traces rest-frame UV light. 

In Figure \ref{sersic_comps}, we plot the $n$ distributions of galaxies and AGNs in the GOODS-S/FIREWORKS catalog, 
for which consistent fits using the same setup are available in both H- and z-bands (see Section 2.1.1). A quick inspection of the top row of panels in this 
Figure and in Figure \ref{galfit_distributions} will show that the basic H-band $n$ distributions and trends are qualitatively preserved
whether one is considering the complete AGN sample or just this subset in GOODS-S. Comparing
the two rows of panels in Figure \ref{sersic_comps}, we see that, while the overall distributions for both AGNs and control are slightly
diskier in the z-band, the AGNs do not show more concentrated light profiles in these bands, as one would have expected from prominent blue
central point sources. In particular, the AGNs appear to be diskier than the control at $0.5<z<1.0$, not bulgier. 
Also both AGNs and inactive galaxies show strongly disky light profiles at $z>1.5$, arguably closer to that of the control
in the rest-frame UV (z-band) than in the rest-frame optical (H-band). The greater similarity of the AGN and control z-band $n$ 
distributions at $z>1.5$ suggests that the excess light in the AGN hosts is at least as red or redder than the typical colour of the control galaxies, 
since it only becomes prominent at optical wavelengths. Therefore, if nuclear light is the major source of the central excess,
it must be reddened considerably by dust along the line-of-sight to the nucleus. However, this will also act to preferentially
extinguish the light from the nucleus, decreasing the contrast of nuclear emission with respect to the host galaxy.

\begin{figure*}[]
\includegraphics[width=\textwidth]{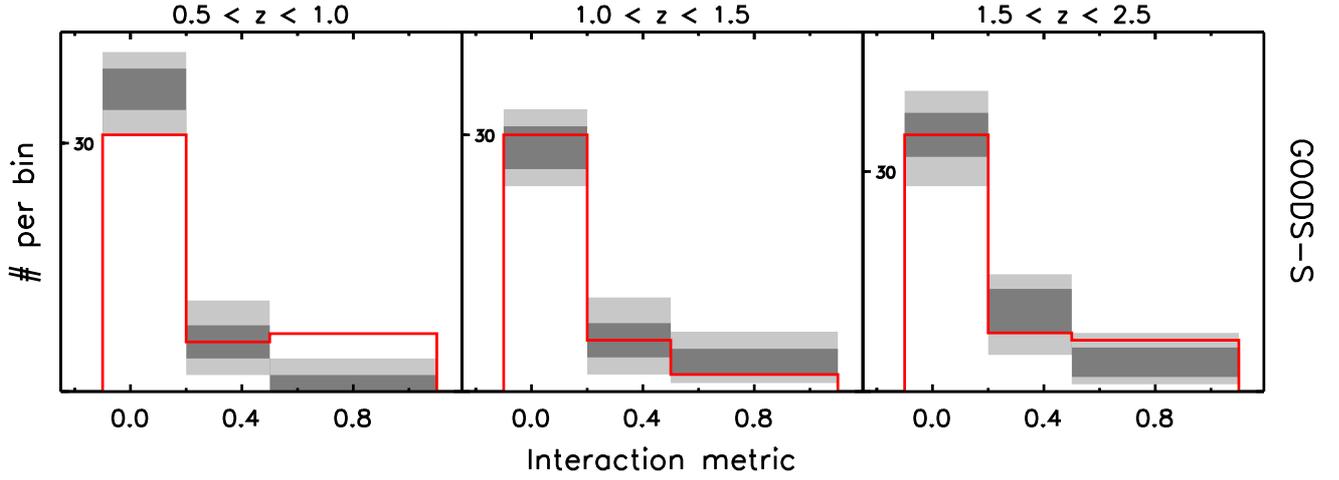}
\caption{Distributions of the visual Interaction Metric (IM) for AGNs and inactive control galaxies.
Panels left to right span three distinct bins in redshift. 1000 draws of a mass-matched control sample are analysed to determine 
the $1\sigma$/$2\sigma$ scatter in the IM distributions for inactive galaxies, shown as dark/light
grey zones in the histograms. Red open histograms show the parameter distributions for the AGNs.
Interacting and merging systems are a minor fraction of galaxies and AGNs at all redshifts. AGNs
are generally equally likely to be in an interacting host as inactive galaxies, though a small excess of
AGNs in mergers may be discernible at low redshifts.
}
\label{im_dists}
\end{figure*}

\begin{figure}[t]
\includegraphics[width=\columnwidth]{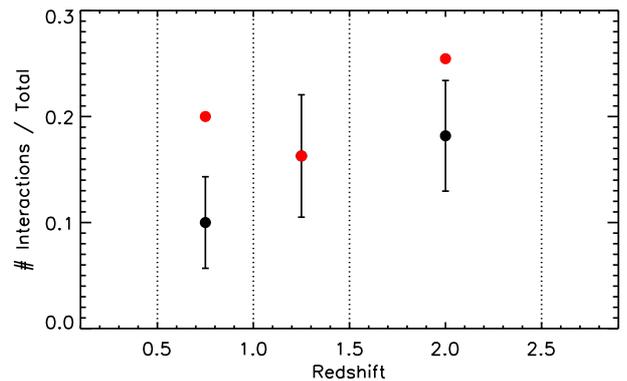}
\caption{Comparison of the observed fractions of AGNs (red points) and inactive galaxies (black points) that are classified as interacting/merging
using the multi-feature approach, in the three fiducial redshift bins (edges shown with dotted lines). The inactive galaxy
fraction is the median value from 1000 draws of a mass-matched control sample to the AGNs in that redshift bin, while
the error bars show the $1\sigma$ scatter from those draws. 
Therefore, the AGNs have an enhancement in the observed interaction fraction over inactive galaxies
at $z<1$ (by $\approx 2 \sigma$), but are comparable to inactive galaxies at $z\sim2$.
}
\label{merg_fracs}
\end{figure}

We explore the consistency of the notion that wide-spread reddened AGN light can influence the centres of the more luminous host
galaxies at $z\sim2$ using Figure \ref{spectral_checks}, in which we contrast a typical galaxy spectrum to AGN spectra with varying
degrees of extinction. The galaxy template shown in the Figure is a typical `Sa' galaxy spectral type from the SWIRE
template library \citep{polletta07}, normalised to a luminosity at $z=2$ that corresponds to an observed frame magnitude of
$H = 24$. The AGN spectrum comes from a composite HST/UV spectrum of radio-quiet QSOs \citep{telfer02} stitched to a 
composite optical spectrum from the Sloan Digital Sky Survey \citep{vandenberk01} (the tabulated spectrum is available at 
http://www.pha.jhu.edu/~rt19/composite/). Since it is empirical, this base spectrum already includes a mild degree of extinction.
The QSO spectrum is normalised to the characteristic luminosity of an AGN with
\lhard$=10^{43.5}$ \ergs, using the X-ray to optical relation for Type 1 AGNs from \citep{lusso10}. 
Different levels of extinction applied to the AGN spectrum yield the spectra shown with dashed lines.
An extinction law from \citep{calzetti94} was assumed, but the conclusions are not strongly dependent on this choice, or
on the choice of the galaxy template used in this exercise.

An X-ray AGN at $z=2$ with the typical luminosity of those in our sample can produce roughly as much light at 5000 \AA\ as a galaxy
at the faint limit considered in this structural study ($H=24$). However, even modest amounts of extinction quickly weaken the relative
blue light of such an AGN, leaving the light from the host galaxy dominant. For an AGN to simultaneously account for
more than 10\%\ of the total light of a system yet be as red as a galaxy spectrum, it would have to be considerably more luminous than 
\lhard$=10^{44}$ \ergs, and only a handful of such sources are found in our AGN sample. This analysis does not preclude
that these rather low luminosity systems could have a much lower X-ray/optical luminosity ratio ($\alpha_{OX}$)
than calibrations based on bright QSOs \citep{vignali03,lusso10} or a much redder intrinsic spectrum
than those of QSOs, such as those used to construct the template shown in Figure \ref{spectral_checks}. Most studies, however,
find a higher $\alpha_{OX}$ in low-luminosity AGN \citep[e.g.,][]{steffen06}. 

All together, these tests lend some support to the notion that high redshift AGNs have a more prominent bulge than equally massive inactive galaxies,
and that the more luminous AGNs are among the bulgiest. We develop the implications of these results in Section 5.




\begin{figure*}[t]
\includegraphics[width=\textwidth]{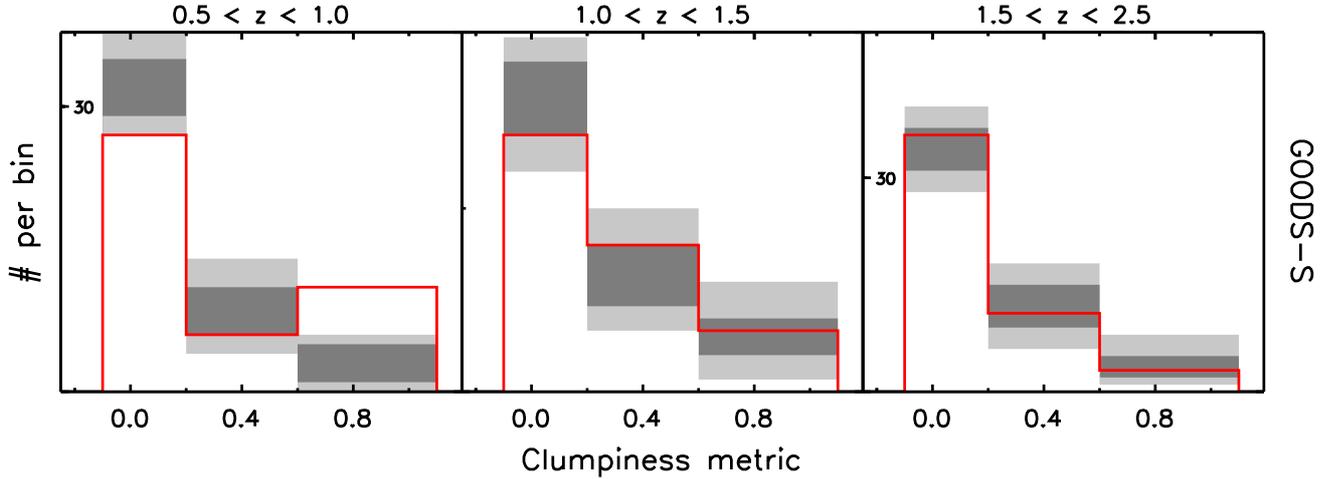}
\caption{Distributions of the visual Clumpiness Metric (CM) for AGNs and inactive control galaxies.
Panels left to right span three distinct bins in redshift. 1000 draws of a mass-matched control sample are analysed to determine 
the $1\sigma$/$2\sigma$ scatter in the CM distributions for inactive galaxies, shown as dark/light
grey zones in the histograms. Red open histograms show the parameter distributions for the AGNs.
Visually clumpy systems are a minor fraction of galaxies and AGNs at all redshifts. At $0.5<z<1.0$,
AGNs are significantly more likely to be in clumpy galaxies. By $1.5<z<2.5$, the CM distributions
are similar.
}

\label{clmp_dists}
\end{figure*}

\subsection{Distributions of the Interaction Metric}

A comparison of the distributions of the visual interaction metric (IM) allows us to quantify
the fraction of AGNs in interacting or merging systems relative to inactive galaxies and
explore the relationship between galaxy interactions and the triggering of nuclear activity.
In Figure \ref{im_dists}, we plot the statistical distributions of IM for inactive galaxies, 
divided coarsely into the three bins delineating ``isolated", ``interacting" and ``merging" systems (Section 2.1.2).
The histograms come from 1000 draws of mass-matched control samples in GOODS-S.
These may be compared to the IM distributions of the CDF-S AGNs (red histograms) in the three fiducial redshift bins. 

A key point to notice is that the majority of sources throughout the redshift range $0.5-2.5$
lie in ``isolated" systems, with IM$=[0.0,0.2]$. The fraction of galaxies classified as ``interacting" or ``merging"
(IM$=[0.2,1.0]$) appears to evolve with redshift (from 18\%\ at $0.5<z<1.0$
to 30\%\ at $1.5<z<2.5$, estimated directly from the control samples). However,
as noted before, morphological K-corrections makes such evolution potentially
hard to interpret.

From Figure \ref{im_dists}, we find a significant difference in the interaction properties
of AGNs and inactive galaxies only at $z<1$. In the low redshift bin, there is an excess of ``merging" systems 
among AGNs (several $\sigma$) at the expense of a smaller number of isolated systems. In both
the higher redshift bins, the IM distributions of AGNs and inactive galaxies are consistent within
the scatter.

Another approach to test the relevance of mergers is to consider the fraction of AGNs and inactive galaxies in the "multi-feature" 
subset of merging/interacting galaxies, described in Section 2.1.2. In Figure \ref{merg_fracs}, we plot
as black points the median fractions of the mass-matched inactive control sample from 1000 draws in each of
the three redshift bins. The errors on these fractions represent the scatter among inactive galaxies from
the draws. The observed fraction of interactions among the AGNs are shown as red points, and may be compared
to the median and scatter of the interaction fractions among inactive galaxies.
In parallel to what we find in the IM distributions, the AGNs
show a significant excess of interacting systems at $z<1$, but are consistent within the scatter with
the interaction fractions of inactive galaxies at $z\sim2$.

\begin{figure}[t]
\includegraphics[width=\columnwidth]{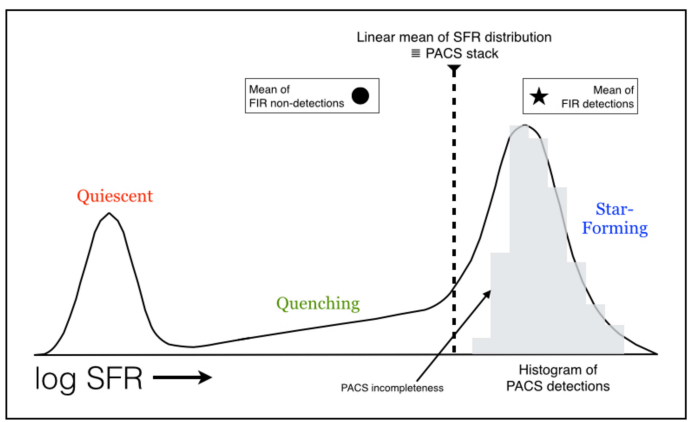}
\caption{An illustration useful for understanding Figures \ref{sfr_sersic}, \ref{sfr_im}, \ref{sfr_clmp}. The solid curve shows a 
qualitative schematic SFR distribution of a population of galaxies, for example, massive galaxies 
at $z\sim2$ where SF galaxies outnumber the quiescent population. It is assumed that the \lfir\ distribution 
is identical and tracks the SFR. The measurable \lfir\ (or SFR) distribution of galaxies individually detected in the
Herschel/PACS maps is shown as a representative grey histogram. At these redshifts, the 
PEP+GOODS-Herschel PACS catalogs are sensitive to FIR luminosities that trace just deeper than the 
ridgeline of the SF Sequence \citep[e.g.][]{rosario13b}, but the completeness of the catalogs at this limit is low ($\approx 30$\%) 
leading to a shallow cutoff in the histogram at the faint end. The mean \lfir\ of the entire
population comes from stacks into the PACS maps (see Section 2.2) and is shown as a dashed vertical line. 
Since stacking is a purely linear process, the mean is shifted towards the star-forming end in this diagram. 
Additionally, the mean \lfir\ of both PACS detected and undetected galaxies can be determined, shown as a star
and circle point as in the later Figures. The schematic also reveals how the fraction of galaxies that are FIR-detected
is closely related to the fraction of moderate and strongly SF galaxies in the population.}
\label{figure_key}
\end{figure}

\subsection{Distributions of Clumpiness Metric}

A measure of the level of disturbance in a galaxy is the clumpiness metric (CM), which is based on the
visual prominence of clumps in galaxies. Galaxy simulations suggest
that the torques driven by large clumps can increase the inflow of gas onto SMBHs, triggering AGN activity \citep{bournaud11, gabor13}.
X-ray stacking studies of clumpy galaxies at $z\sim1$ support this theoretical insight \citep{bournaud12}.

From 1000 draws of mass-matched control
samples, we determine the statistical distributions of CM for inactive galaxies in our fiducial redshift bins, as shown 
in Figure \ref{clmp_dists}. There is a sharp dropoff in the distribution with CM; most galaxies only show low levels of visual
clumpiness. The fraction of clumpy systems (defined here as CM $>0.2$) increases mildly with 
redshift from 25\%\ at $0.5<z<1.0$ to 32\%\ $1.5<z<2.5$. This is expected
given the current understanding of galaxy evolution where more turbulent high redshift galaxies contain
more clumpy disks. However, since the bluest band (F606W) used for the visual 
assessment of clumpiness changes from rest-frame $\sim3500$\AA\ in the
low redshift bin to $\sim2000$\AA\ in the high redshift bin, a morphological k-correction could also play
some role in this apparent evolution.

The distribution of clumpiness in AGNs is overplotted in the Figure as a red solid line. In many
ways, the trends found in IM in Figure \ref{im_dists} are also reflected in CM.
At $0.5<z<1.0$, AGNs are more likely to be found in clumpy galaxies, at a significance of several times the scatter. At these
redshifts, we estimate a clumpy fraction among AGNs of $39^{+7}_{-7}$\%.
By $1.5<z<2.5$, the fraction of clumpy AGNs drops to $28^{+7}_{-6}$\%, though it is formally consistent with the
clumpy fraction of AGNs all redshifts. This fraction is comparable to the clumpy fraction of inactive
galaxies at the same redshifts, and an examination of the right panel in Figure \ref{clmp_dists} also shows that AGNs
and inactive galaxies have similar CM distributions at higher z. This may be  contrasted with their rather different
distributions at low redshift. Broadly, there appears to be a preference for the clumpy fraction of
AGNs to change less slowly with redshift, akin to their light profile parameter distributions.
This implies that the characteristic structure of AGN hosts may remain approximately invariant 
with redshift between $z=2.5$ and $z=0.5$.

\section{SFR--structure relationships}

 Valuable insight into the scenarios that link AGN fueling and host galaxy properties comes from combining
 information about SFR and host structure. In this section, we explore these relationships using the deep Herschel/PACS
 FIR data in the two GOODS fields. We employ two main types of measurements in this analysis. Firstly,
 we use mean FIR luminosities of AGNs and inactive control galaxies. These are estimated directly from PACS fluxes
 for the subset of sources detected in the PACS maps, and from stacks on the residual maps for the PACS-undetected sources (Section 2.2).
 Secondly, we employ the FIR detection fraction, defined as the fraction of sources detected in both the 100 and 160 \mics\ 
 PACS maps. The PEP+GOODS-Herschel photometry detects massive galaxies that lie on
 or just below the SF sequence out to $z=2.5$ \citep{rosario13b, magnelli13}. Therefore, the
 FIR detection fraction among a certain population of galaxies is a measure of the fraction 
 of moderate and strongly star-forming galaxies in that population. In absolute terms, this measure is sensitive to 
 the depth and completeness of the PACS catalogs, and the redshift-dependent FIR luminosity function.
 We only use it here as comparative tool to understand differences between AGNs and inactive galaxies, and
 specifically warn against the over-interpretation of any trends with redshift.

\begin{figure*}[h]
\centering
\includegraphics[width=0.9\textwidth]{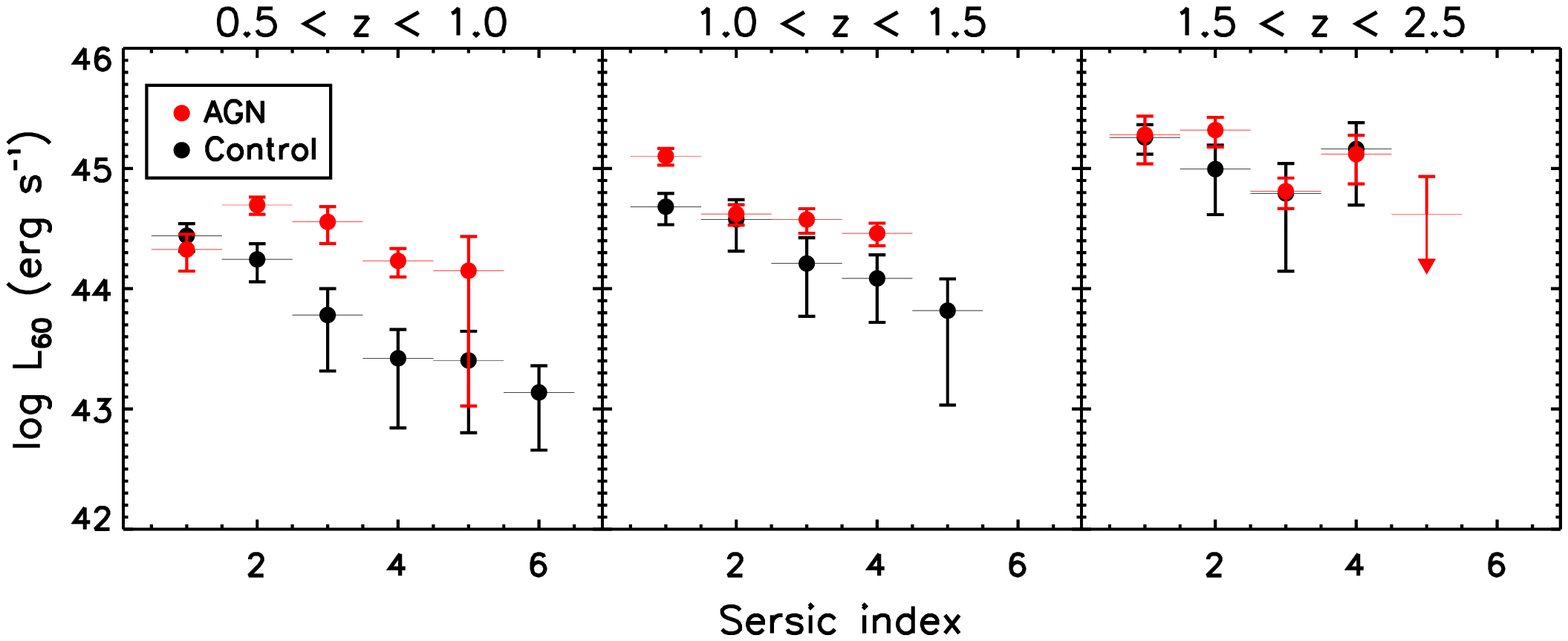}
\includegraphics[width=0.9\textwidth]{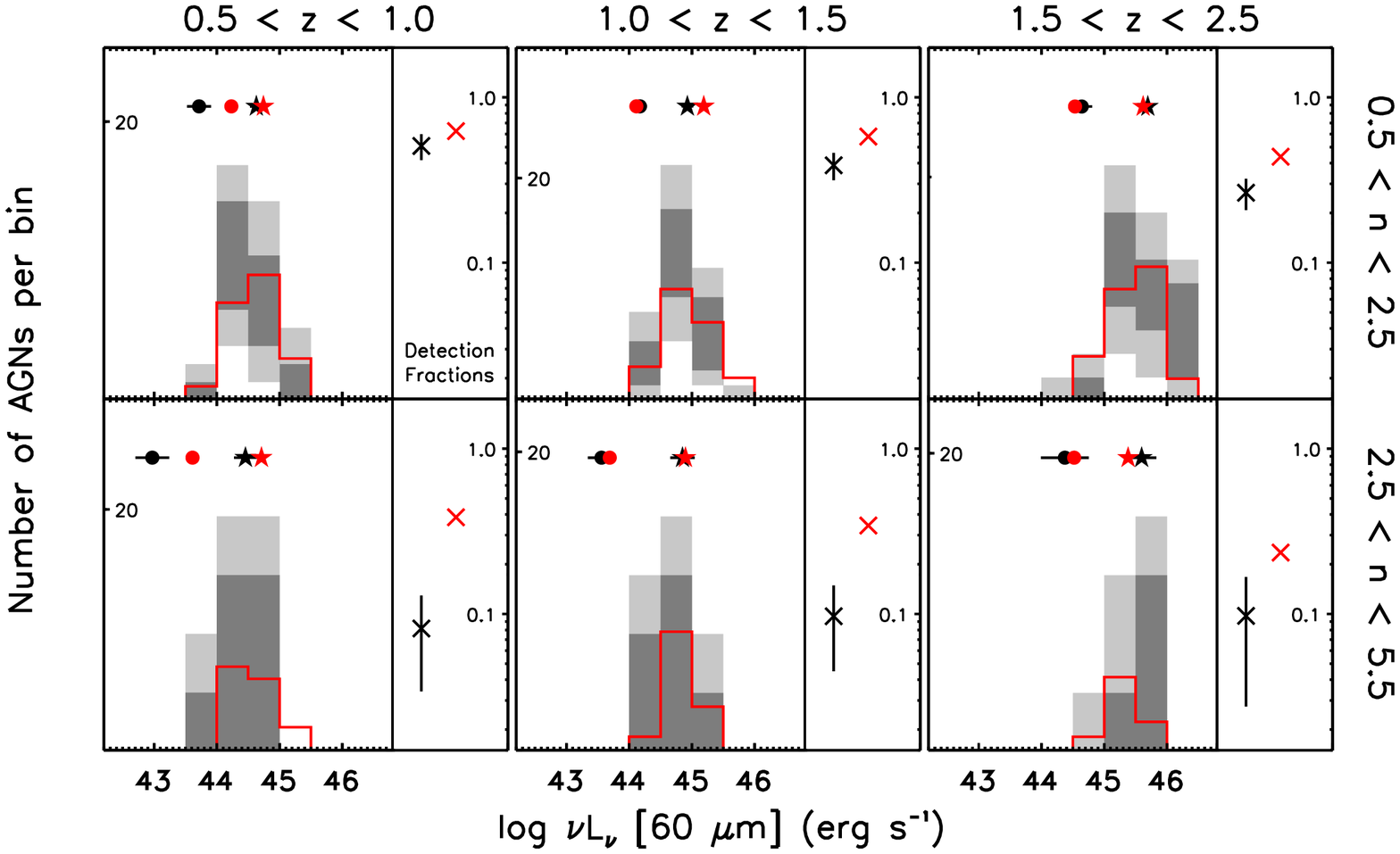}
\caption{Mean 60 \mics\ monochromatic luminosities (\lfir) and luminosity
distributions of AGNs and inactive control galaxies in both GOODS fields combined, as a function of H-band
\sersic\ index ($n$). Panels left to right span three distinct bins in redshift.
In the top row, the mean \lfir\ from combinations of
detections and stacks are compared in bins of $n$. The X-axis error bars show
the extent of these bins. The errors on the AGN measurements (red points) 
are determined from bootstrapping into the AGN subsample in each bin in redshift and $n$. 
The errors on the inactive galaxy measurements (black points)
come from the analysis of an ensemble of mean \lfir\  from 100 random draws of 
mass-matched control galaxies. In the lower two rows, each panel is split into two subpanels.
\lfir\ distributions and PACS detection fractions
for AGNs and inactive galaxies are plotted in the left subpanels for two coarse bins in \sersic\ index.
Histograms show the \lfir\ distributions of PACS-detected AGNs and inactive galaxies. 
The 100 random draws of mass-matched control galaxies
are used to determine the $1\sigma$/$2\sigma$ scatter in \lfir\ for inactive galaxies, 
shown as dark/light grey zones in the histograms. The red histograms are the \lfir\ distributions for AGNs.
The mean \lfir\ corresponding to these histograms are plotted for PACS-detected AGNs as red star points and for
inactive galaxies as black star points. Error bars on the latter are the rms scatter of the mean of the control
samples. The mean \lfir\ of PACS-undetected AGNs (red circle points) and inactive galaxies (black circle points)
are also compared similarly. PACS detection fractions are shown in the right subpanels. 
Red cross points denote AGNs and black cross points denote inactive galaxies, with rms scatter shown as vertical
error bars on the latter points. See Section 4.1 for a discussion.}
\label{sfr_sersic}
\end{figure*}
\begin{figure*}[h]
\centering
\includegraphics[width=0.9\textwidth]{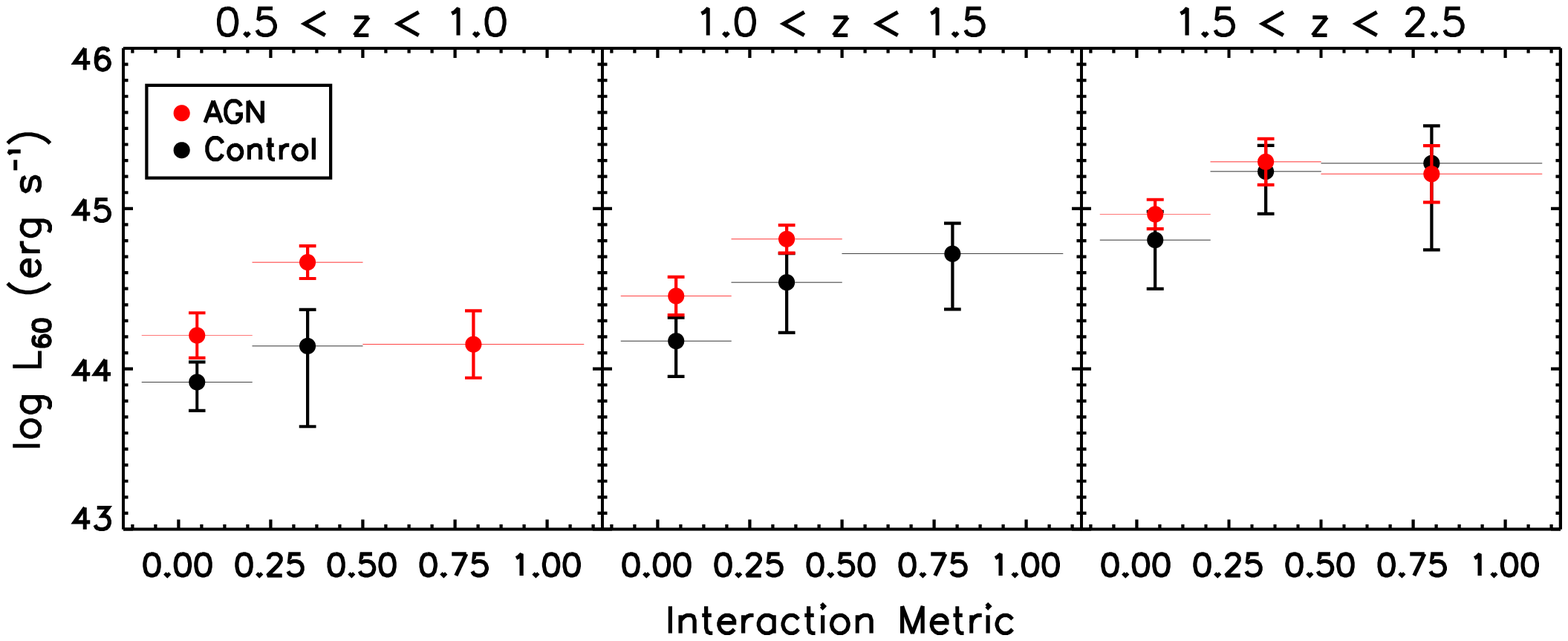}
\includegraphics[width=0.9\textwidth]{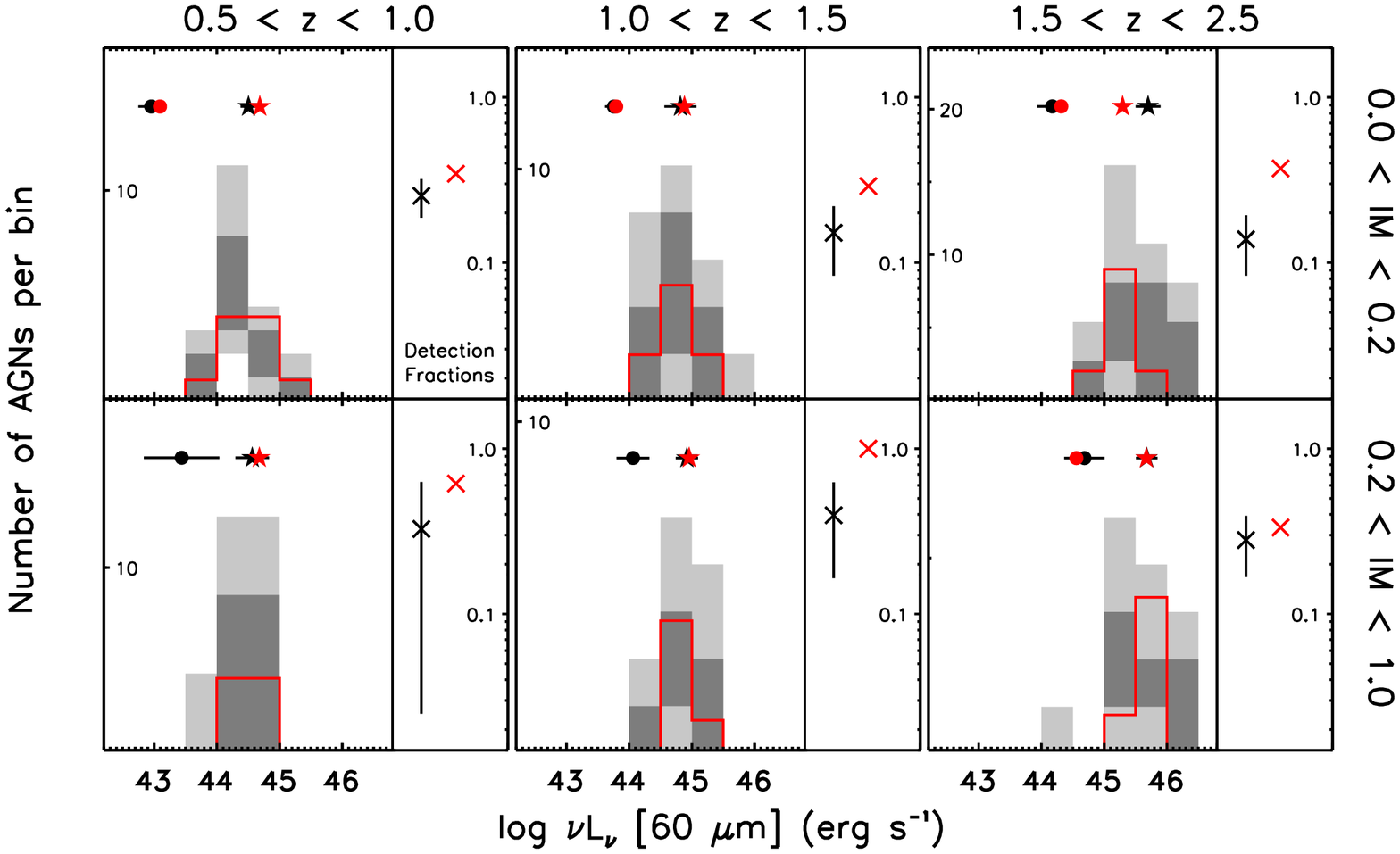}
\caption{Mean 60 \mics\ monochromatic luminosities (\lfir) and luminosity
distributions of AGNs and inactive control galaxies in GOODS-S, as a function of visual
Interaction Metric (IM). Panels left to right span three distinct bins in redshift.
In the top row, the mean \lfir\ from combinations of
detections and stacks are compared in bins of IM. 
In the lower two rows, each panel is split into two subpanels.
\lfir\ distributions and PACS detection fractions
for AGNs and inactive galaxies are plotted in the left subpanels for two coarse bins in IM.
PACS detection fractions are shown in the right subpanels. 
Details of the plot are identical to those of Figure \ref{sfr_sersic}, except for a difference in
the structural measure. See Section 4.2 for a discussion.}
\label{sfr_im}
\end{figure*}
\begin{figure*}[h]
\centering
\includegraphics[width=0.9\textwidth]{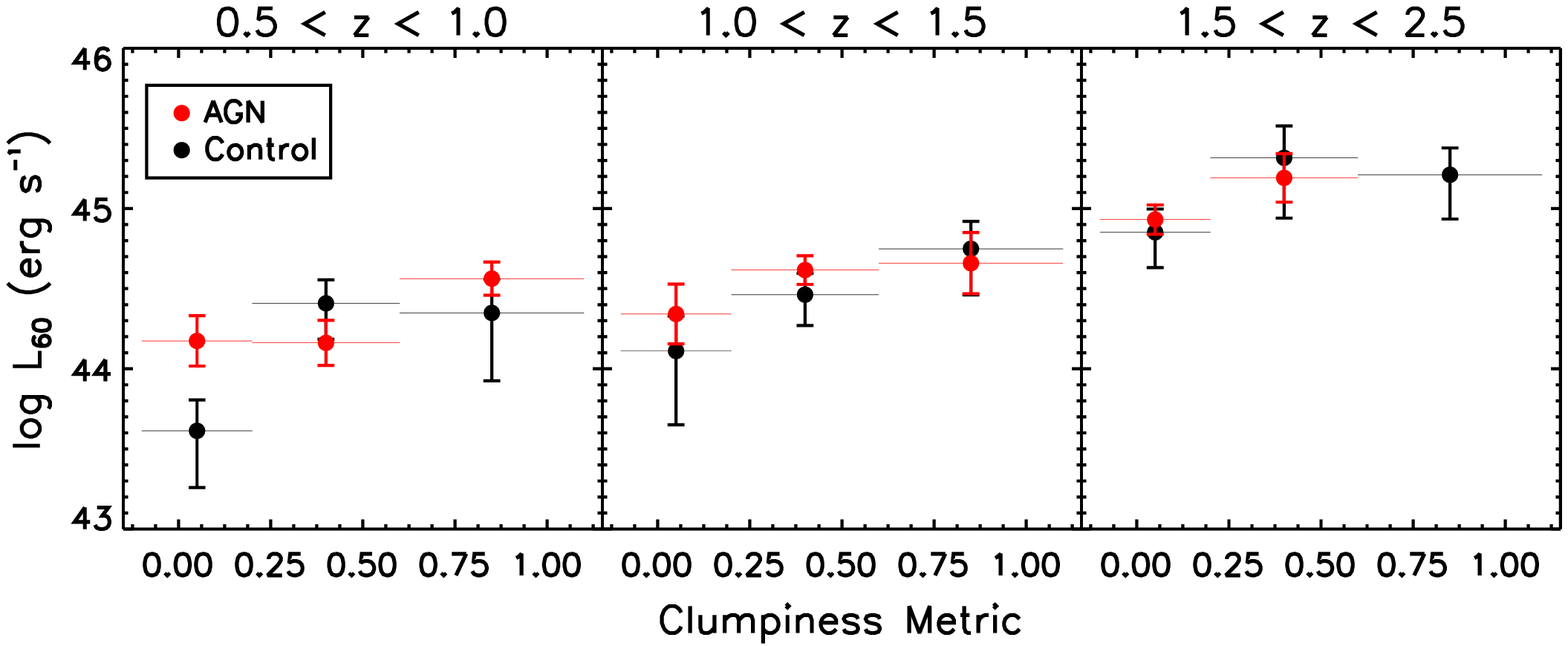}
\includegraphics[width=0.9\textwidth]{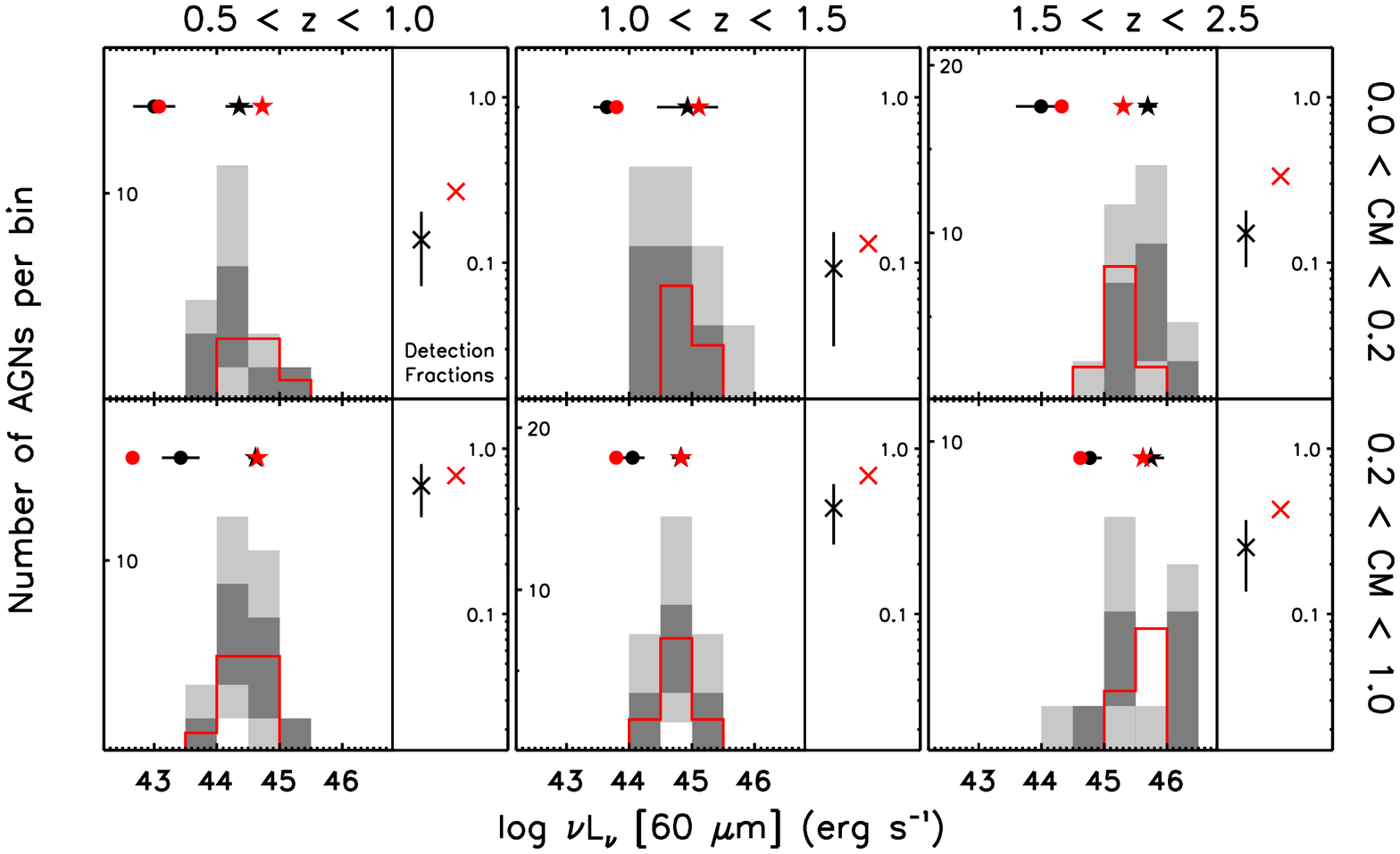}
\caption{Mean 60 \mics\ monochromatic luminosities (\lfir) and luminosity
distributions of AGNs and inactive control galaxies in GOODS-S, as a function of visual
Clumpiness Metric (CM). Panels left to right span three distinct bins in redshift.
In the top row, the mean \lfir\ from combinations of
detections and stacks are compared in bins of CM. 
In the lower two rows, each panel is split into two subpanels.
\lfir\ distributions and PACS detection fractions
for AGNs and inactive galaxies are plotted in the left subpanels for two coarse bins in CM.
PACS detection fractions are shown in the right subpanels. 
Details of the plot are identical to those of Figure \ref{sfr_sersic}, except for a difference in
the structural measure.
See Section 4.3 for a discussion.}
\label{sfr_clmp}
\end{figure*}

\subsection{SFR as a function of \sersic\ index}

 
In Figure \ref{sfr_sersic}, we compare, in the upper row of panels, the mean \lfir\ of AGNs (red points) and mass-matched
control galaxies (black points) in bins of redshift and \sersic\ index $n$. The errors on the mean \lfir\ for AGNs come
from a bootstrap resampling of the AGN sample itself (Section 2.2), while the errors on the mean \lfir\ for the inactive galaxies
are determined from 100 draws of mass-matched controls. 

For the control sample, we find a clear and consistent drop in the mean \lfir\ with $n$ in the low and
intermediate redshift bins, although the trend appears to flatten out at $1.5<z<2.5$. This is as expected - most of
the SF in the local Universe and at higher redshifts is associated with galaxies having substantial disks \citep[e.g.][]{kauffmann03a,wuyts11,lee13},
although at $z\sim2$, a sizeable population of blue star-forming compact spheroids is also found \citep{barro13}. In contrast
to the inactive galaxies, the AGNs show a much flatter dependence of mean \lfir\ with $n$, such that low \sersic\
galaxies and AGNs have comparable FIR luminosities, but, especially at $z<1.5$, high \sersic\ AGNs are
significantly more luminous than high \sersic\ galaxies. In essence, this means that the enhancement in mean SFR
among AGNs is more pronounced in bulgy or early-type galaxies.

In the lower panels, we compare the \lfir\ distributions of AGNs and inactive galaxies, splitting them by redshift 
and coarse bins in \sersic\ index: $0.5<n<2.5$ cover disk-like exponential light profiles, while $2.5<n<5.5$ include
bulge-like De Vaucouleurs profiles. The histograms show the \lfir\ distributions of PACS-detected sources. As before, 
the dark/light grey shaded regions show the 1$\sigma$/2$\sigma$ scatter of the mass-matched control sample, and
the AGNs are represented by colored open histograms. The mean \lfir\ of the PACS-detected objects are shown
as star points, where error bars represent one standard deviation in the mean \lfir\ of the control sample. 
The \lfir\ distributions of PACS-detected AGNs are statistically very similar to those of the control sample, 
except among isolated galaxies at high redshift, where the AGNs show a slightly
lower fraction of FIR bright systems. 

We also compare the mean \lfir\ of PACS-undetected sources based on stacks, which are shown in Figure \ref{sfr_sersic}
as large circle points. Again, where measurable, the AGNs have comparable mean FIR luminosities to the
inactive galaxies, except in the low redshift bin, where they appear to be more luminous by a factor of a few. Finally, we
show the PACS detection fraction of AGNs and inactive control galaxies as cross points in the insets to each
panel of the lower Figure. As has been noted by us in earlier studies, AGNs as a whole are significantly more likely to be detected in PACS
than equally massive inactive galaxies, suggesting that nuclear activity prefers SF hosts \citep{santini12, rosario13b}.
The biggest consistent difference between AGNs and control galaxies can be found in the detection
fractions. For low $n$,  AGNs have a mildly elevated FIR detection fraction over the control, comparable to the full
AGN sample. However, high $n$ AGNs are several times more likely to be detected in PACS than inactive galaxies with similar
structure. This is the likely reason for the enhanced mean \lfir\ seen among bulgy AGNs in the upper panels.

\subsection{SFR as a function of Interaction Metric}

Our analysis of Section 3.4 suggests that merger rates among AGNs at $z<1$ are enhanced.
If substantial fueling of AGNs and bursts of SF are both instigated by gas-rich mergers, 
nuclear activity is expected to be preferentially found in those interacting systems that also exhibit strong SF.
We test this notion in GOODS-S using the visual interaction metric to separate isolated, interacting and merging systems. 

In Figure \ref{sfr_im}, we compare, in the upper row of panels, the mean \lfir\ of AGNs (red points) and mass-matched
control galaxies (black points) in bins of redshift and IM. 
In general, we find that the FIR luminosities are not strongly dependent on IM for both classes of sources. There is a
rise between ``isolated" and ``merging" systems, by a factor of $\sim 0.5$ dex in the two higher redshift bins. 
The mean \lfir\ of the AGNs is somewhat enhanced with respect to inactive galaxies at $z<1.5$, by $\approx 0.3$ dex. This is broadly consistent
with the level of FIR enhancement found for all X-ray AGNs in \citet{santini12}. There is a hint that the enhancement may be higher
among ``interacting" galaxies in the lowest redshift bin, but this pattern is not observed at intermediate redshift. At $1.5<z<2.5$,
AGNs and inactive galaxies show similar mean \lfir.

In the lower panels, we compare the \lfir\ distributions of AGNs and inactive galaxies divided by redshift and into bins of
$0.0<$IM$<0.2$ (isolated galaxies) and $0.2<$IM$<1.0$ (all interacting and merging systems for sufficient statistics). 
The \lfir\ distributions of PACS-detected AGNs are generally indistinguishable 
from those of the control sample, except among isolated galaxies at high redshift, where the AGNs show a slightly
lower fraction of FIR bright systems. The mean \lfir\ values (star points) are consistent with the visual 
appearance of the distributions and also indicate the significance of the low mean \lfir\ among AGNs detected by PACS at $1.5<z<2.5$.

PACS-undetected AGNs and inactive galaxies also have comparable mean FIR luminosities. 
From the detection rates comparison in the inset panels, the general detection rate of AGNs is significantly higher than that
of inactive galaxies among both isolated and interacting systems. However, the difference in detection 
rates between AGNs and inactive galaxies is not clearly dependent on the IM of the hosts, 
suggesting that starbursts and nuclear activity are not any more coevally connected in mergers than in isolated galaxies.


\subsection{SFR as a function of Clumpiness Metric}

We have shown in Section 3.5 that AGNs are more likely to be in clumpy galaxies, at least
at $z<1.5$. In Figure \ref{sfr_clmp}, we explore trends between FIR luminosity and the visual clumpiness of galaxies. As
before, we look first at the mean \lfir\ of AGNs and the inactive control sample in bins of redshift
and CM (top row of the Figure). Firstly, there is a trend towards mildly elevated \lfir\ among clumpy galaxies at all redshifts,
which is expected since clumps are a signature of unstable star-forming disks \citep{dekel09, ceverino10}.
At $z<1.5$, the enhancement in mean \lfir\ for AGNs is small or non-existent among clumpy systems, but appears to be
larger for smooth systems. The differences between the mean \lfir\ of AGNs and inactive
galaxies as a function of CM reduces and possibly goes away by $z\sim2$.

The two lower rows of Figure \ref{sfr_clmp} compare \lfir\ distributions of the two populations. To allow sufficient
number statistics, we use coarser bins in CM than in the upper panel: $0.0<$IM$<0.3$ indicates smooth galaxies, 
while $0.3<$IM$<1.0$ includes both mildly and strongly clumped galaxies. AGNs and inactive galaxies show fairly
similar distributions in \lfir\ in both CM bins, both among PACS-detected and -undetected systems. In the high redshift bin, the
AGNs are mildly weaker in mean star-forming properties compared to the inactive population.
A comparison of PACS detection fractions shows again that AGNs tend to be detected a little more
often than the control sample. There is a mild indication that the differences in detection rates are smaller among clumpy systems
than among smooth systems.

\section{Discussion}

\subsection{Differences in structure and star-formation between AGNs and inactive galaxies}

In earlier sections, we analyzed structural and SFR patterns of X-ray selected AGN host galaxies 
and compared them to equally massive inactive galaxies. 
Broadly, AGNs are structurally similar to non-AGNs, whether in terms of galaxy light profiles,
clumpiness or the incidence of interactions/mergers. Both sets of galaxies are primarily isolated, smooth and moderately bulgy systems.
AGNs of all structural categories are more likely to detected in the FIR, 
although their SFR distributions, tracked using \lfir\ distributions and stacks, are not vastly different
from inactive galaxies \citep[a more extensive treatment can be found in][]{rosario13b}. Despite this, we do uncover some significant 
differences through the use of a careful
comparison of statistical distributions. 

First, AGNs show little systematic variation in their rest-frame optical light profiles despite the obvious changes in the profiles of non-AGNs
with redshift. At $z\sim1$, massive inactive galaxies are primarily bulgy, with a typical \sersic\ index
of 3, similar to the AGNs. By $z\sim2$, inactive galaxies become mostly disk-like and show 
\sersic\ indices peaking at $n=1$. In contrast, the typical \sersic\ index
of AGNs changes little between $z=2.5$ and $z=0.5$. Comparisons of axis ratio distributions also
highlight the rounder profiles of AGN hosts at $z\sim2$.

Comparing H-band and z-band light profile distributions reveals that these differences are produced by a
central light excess in AGNs that is at least as red as the rest of the galaxy. Rest-frame UV
light profiles of AGNs and non-AGNs at these redshifts are similar to their rest-frame optical
light profiles. This suggests that the star-forming disk component, 
which likely dominates the rest-frame UV light, is comparable in shape in both populations of galaxies, whereas
the central red excess is more pronounced in AGNs. At $z>1$, the higher \sersic\ AGN hosts are also more
luminous, pointing to a possible relationship between the origin of the light excess and the nuclear luminosity.
While this excess could arise from reddened nuclear emission
from an obscured active nucleus, the low luminosity of most X-ray selected AGNs in the CDFs 
implies instead that a more prominent stellar bulge may exist in AGN hosts \citep[Section 3.3.1 and][]{rosario13a}.
Noting the important caveats in this conclusion outlined in Section 3.3.1, we proceed with the cautious
implication of our results that the relationship between bulges and the existence of super-massive black holes 
was already in place at $z\sim2$, although we refrain from any speculation about the nature of that relationship 
or its evolution based on our present analysis. In the discussion on AGN fueling modes (Section 5.3), 
we will consider the implications of both alternatives: AGNs are in more bulge-dominant hosts at $z\sim2$, or
AGNs are in structurally similar hosts to inactive massive galaxies at $z\sim2$.

While the light profiles of AGNs are similar to inactive galaxies at $z\sim1$ (and probably not affected by AGN contamination), 
their SFRs show different trends with \sersic\ index. In particular, AGNs do not share the strong drop of SFR with $n$ characteristic of inactive
galaxies. An examination of the SFR distributions reveal that these high \sersic\ AGN hosts show both higher
SFRs and a much higher FIR detection rate. This suggests that most bulge-dominated AGN hosts
are indeed forming stars, while most bulge-dominated inactive galaxies are fairly quiescent. At $z\sim2$, high \sersic\
AGNs still maintain high detection rates over inactive galaxies, but the differences in the mean SFRs are less
pronounced, probably because all massive galaxies, regardless of structure, show larger SFRs at these redshifts.

Using the visual clumpiness metric (CM), we show that 
the majority of massive galaxies and AGN hosts at all redshifts are in relatively smooth galaxies.
Despite this, we find a significant preference for AGNs to lie in clumpy hosts at
$z \lesssim 1$. This preference weakens towards $z\sim2$, while the relative fraction of clumpy galaxies, both active
and inactive, increases steadily. If the presence of large clumps is related to the level of turbulence or 
instability in galactic gaseous disks, then the AGNs and inactive galaxies at $z\sim2$ show the same level
of such disturbance, while at $z<1$, AGNs appear to prefer clumpier galaxies and are more likely
to be found in turbulent hosts.

Across redshift, both active and inactive clumpy galaxies have mutually comparable SFRs. In addition, at low redshifts AGNs
in smooth hosts display significantly enhanced SFRs. Note that these smooth hosts include most of the high \sersic\
AGNs as well. Therefore, the large enhancement in SFR observed among both bulgy and smooth AGN hosts are
two sides of the same coin, tracing much of the same population of host galaxies.

Additionally, our study allows us to compare the relative incidence of interacting or merging systems in AGNs
and equally massive inactive galaxies, using the visual interaction metric (IM) or the ``multi-feature" merger subset. 
At $z<1$, both tracers suggest a significant enhancement of mergers among AGN hosts.
The fraction of interacting systems among AGNs in the $0.5<z<1.0$ redshift bin is quite consistent with the fractions found 
for local hard X-ray selected Swift/BAT AGNs \citep[18\% -- 25\%;][]{koss10}.
Towards $z\sim2$, AGNs and inactive galaxies have essentially identical IM distributions. Therefore, the 
enhancement in AGN activity in mergers seen in local and low redshift appears to become less pronounced at higher redshifts.

Across IM, AGNs at $z\sim1$ typically show substantially higher mean SFRs than the control sample. These differences
are driven by a combination of mildly higher individual SFRs and a consistently higher FIR detection rate. The
enhancement in AGNs is consistent with the notion that nuclear activity is generally associated with gas-rich galaxies
\citep{santini12, rosario13b}.  We do not discern any clear indication that the SFRs 
of interacting galaxies hosting AGNs are additionally boosted over isolated galaxies
hosting AGNs, although this is predicted by most models of co-eval SMBH fueling and starbursts
in galaxy interactions. If there is such a secondary enhancement, it is minor when considering the 
ensemble of visually identified interacting systems. On the other hand, this ensemble includes 
many gas-poor mergers, minor mergers or fly-bys, which do not necessarily conform to the predictions of merger
simulations. A careful treatment of merger samples, with additional classification based on the gas content, 
mass ratios of the merging components, as well as the stage of the interaction, may reveal finer relationships. This
may be possible in future work that uses the entire CANDELS area, but is beyond the statistical capability of our analysis.

\subsection{Comparison with existing studies of X-ray selected AGNs}

A number of studies using various methods have explored the structural properties of AGN hosts at intermediate and 
high redshifts, sometimes using samples that overlap significantly with those used here. One advantage of our
study is a combination of visual and analytic measures of galaxy structure over a wide range in redshift, 
which enables us to compare our results to a broader subset of relevant work from the literature.

We are generally consistent with earlier studies which used analytic measures of AGN host structure, typically at 
$z < 1$ \citep{grogin05, pierce07}. Some of the differences between the AGNs and inactive galaxies in these
earlier studies, such as higher early-type fractions or bluer colors among AGNs, 
are due to the improper choice of control samples, for example through matching by a blue optical luminosity
rather than stellar mass. Such control samples contain a larger number of lower mass blue star-forming galaxies than
can be found among AGN hosts, which biases both structural and SF measures. The need for a stellar mass-matched
comparison sample is critical for a fair assessment of the AGN hosts \citep[e.g.,][]{villforth14}. 

\citet{kocevski12} compared AGNs and mass-matched control galaxies at $1.5<z<2.5$ in CANDELS/CDF-S using 
a visual classification scheme archetypical to the one in this paper. While we do not employ
the visual separation into spheroids and disks as used in \citet{kocevski12}, we have verified, using our
visual classification catalogs, that our results are completely consistent with this earlier work (Appendix B). Among AGNs
at these redshifts, most host galaxies are classified as disks, with a low merger/disturbed fraction and 
only a minor enhancement in the visual spheroidicity over inactive galaxies. At first
glance, this seems to be at odds with the high \sersic\ indices seen among AGNs at $z\sim2$. These differences
stem primarily from the higher sensitivity of visual classifiers to the appearance of disks and a rather weak ability
to discern variations in steepness of light profiles among spheroidal systems. An elongated $n=2.5$ galaxy will be
classified as a disk as easily as an elongated $n=1$ galaxy. Only GALFIT light profile modeling
can adequately reveal the central light excess we find in AGN hosts at $z\sim2$. A similar limitation will apply
to visual studies of lower redshift AGNs from optical images such as \citet{cisternas11}, 
although the higher resolution of the HST/ACS imaging could potentially reveal bulges more effectively. 
Visual methods are considerably more sensitive to distortions, asymmetries and disturbances in galaxies.

From HST/WFC3 early release science imaging in GOODS-S, \citet{schawinski11} determined that AGN hosts
at $1.5<z<3$ were mostly in disky hosts, with typical GALFIT-based \sersic\ indices peaking at $n=1$. This result is at odds
with our finding that AGNs at $1.5<z<2.5$ have typical $n$ of 2.5. One major difference lies in the fact that 
the AGNs in \citet{schawinski11} were fit using a two component galaxy model, with the assumption that the central
light excess in these objects arises due to nuclear contamination. We only fit our galaxies with a single model profile
and our two band studies demonstrated that the central excess is red and likely due to a bulge. By subtracting away
a large part of the central bulge component, \cite{schawinski11} may be lowering the effective \sersic\ indices
in their AGNs (see Appendix A for a demonstration of this effect). To better understand the differences between 
these results, careful multi-component modeling of the WFC3 images of AGNs and inactive galaxies is needed. 
Simulations of the influence of multi-component fits on the recovery of bulge and disk parameters in realistic galaxies is 
needed, akin to the approach taken by \citet{simmons08} and \citet{gabor09} for AGNs imaged with ACS. 

On the topic of merger incidence, the higher merger rates we find among AGNs at $z<1$ is consistent with the
level of enhancement of AGN activity in low redshift mergers \citep{silverman11, sabater13}.
In addition, the weakening of a merger connection at $z\sim2$ is consistent with \cite{kocevski12}.
Our results at $z<1$ are in some tension with the HST/ACS study of XMM-COSMOS AGNs from \cite{cisternas11}, which
found no enhancement in the interaction fraction over the control sample of inactive galaxies in an overlapping redshift
range, despite a similar statistical power to our own study. This may arise due to the higher median luminosities of the XMM-COSMOS
sample, roughly an order of magnitude greater than the CDFs at these redshifts. Alternatively, and more likely, the differences may
stem from the different approaches used to construct a control sample; we match galaxies by stellar mass, 
while \cite{cisternas11} match in redshift and apparent F814W (I-band)
magnitude, correcting for any emission from a nuclear point source. The use of a blue optical rest-frame band to match galaxies
will allow lower mass star-forming galaxies with a low M/L to enter the control, leading to a different stellar mass distribution between
AGNs and the control sample \citep{xue11,rosario13a}. While it is not immediately apparent how this can affect merger incidence,
a careful assessment of matching criteria is required before these differences are to be understood. 

\subsection{Insights into AGN triggering scenarios}

A radiatively luminous AGN, including one bright in the X-rays, is triggered by dense gas falling onto an accretion disk around an SMBH.
A major area of inquiry in the field of active galaxies pertains to how gas gets from scales of the host 
galaxy down to the black hole. Violent processes, such as galaxy mergers, are
very effective at stripping angular momentum from gas, sending it into the centres of galaxies to produce dense
compact structures which can fuel synchronized starbursts and luminous AGNs. However,  secular processes can also
bring gas into the vicinity of the SMBH, as evinced by the gas-rich circumnuclear environments of settled
disk galaxies such as our own Milky Way \citep[e.g.,][]{morris96,kruijssen13}. This gas can amply fuel low and moderate luminosity
AGNs \citep{hopkins06a}, and, depending on the mechanisms relevant for the small-scale inflow of gas around the nucleus, could even
fuel luminous phases such as quasi-stellar objects \citep[e.g.][]{gabor13}. 

The morphology of a galaxy is an aggregate of its evolution over many epochs. Once a galaxy's stars
are redistributed by violent processes into a spheroid, this marker of its merging and inflow history is preserved, even
if further inflow of gas and stars may settle into a later-forming disk. Folding together the morphological information
of an AGN host, as a measure of its integrated evolution, and its SFR, as a measure of its current evolutionary state,
can potentially constrain the importance of various AGN triggering models.

We use the results of our study at $z\lesssim1$ to demonstrate, through a simple heuristic example, the way
structure and SFR may be jointly employed in testing fueling mechanisms. More quantitative and discriminatory 
tests will involve a comparison to observable predictions from cosmological semi-analytic models or hydrodynamic simulations.

AGNs at all redshifts are found in all forms of massive galaxies: those with disks, bulges and pure spheroids; isolated, clumpy, interacting or
merging. Clearly, neither purely violent processes or purely secular inflow can account for all AGN triggering. 
The median \sersic\ index of $\approx 2.5$ implies that the typical AGN host is a disk galaxy with a substantial bulge, 
as is also confirmed by visual and analytic estimates from earlier studies \citep[e.g.][]{grogin05, pierce07, gabor09}.
The prominence of a spheroid or bulge is a signature that violent processes likely played a role in the structural
evolution of these galaxies. But the presence of a cold disk component, as well as the similarity of the \sersic\ index
distributions of active and inactive galaxies, implies that either those violent events directly fuel only a small fraction of 
AGNs, or that such violent mechanisms do not preclude the simultaneous formation or preservation of a galaxy disk
\citep{robertson06c,hopkins09, dekel09, ceverino10}.
The typical SFR of AGN hosts in disk galaxies is comparable to those of inactive galaxies and consistent with that
of the SF Sequence \citep{mullaney12, rosario13b}: AGN activity must be driven in these
galaxies by secular processes that do not strongly disturb the star-forming equilibrium of their gas disks. Therefore, any
consideration for AGN triggering scenarios must be mixed, with both secular and violent components. Theoretical insight
suggests that this mix evolves with redshift in the low and moderate luminosity AGN population \citep{hopkins13}.

We consider a scenario where AGN triggering is not directly linked to processes that govern the larger scale galaxy.
In this view, the role of the outer galaxy and its environment is to simply supply gas to the inner kpc around the 
SMBH, which may arrive either through violent torques and relaxation, or through longer secular means. 
Once there, small-scale physics will govern the final infall of this gas to the accretion disk, modulating the duty
cycle of the active phase. SF indicates the presence of cold gas in a galaxy, which is why we find AGN activity is 
enhanced in star-forming galaxies across almost all galaxy morphologies and forms.
In galaxies with very low gas reservoirs, such as massive ellipticals, the chance
of hosting enough gas in the circumnuclear regions is quite low. Indeed, only the small subpopulation of 
early type galaxies with significant gas content contain an X-ray bright AGN and also display detectable star-formation. 
This explains the preponderance of SF in early-type AGN hosts. Among disk-dominated galaxies, 
the presence of clumps are a signature of higher gas fractions
and more turbulent disks, which is why AGN hosts are mildly more clumpy at $z\lesssim1$.

However, this simple scenario does not adequately explain the higher incidence of mergers among AGNs at these redshifts.
Therefore, a channel must exist which involves a direct connection between merger-driven torques on galaxy scales and gas inflow on
to the SMBH on nuclear scales \citep[e.g.][]{hopkins10}. An potential signature of this direct ``violent" channel is an average
enhancement in the SFR of AGN-hosting mergers, which has been reported in detailed studies of local galaxy pairs and post-merger
remnants \citep{liu11,ellison11, ellison13}. Unfortunately, the number of inactive visually classified mergers in our sample is too
small to obtain a strong FIR detection at $z\lesssim1$, and we cannot test this notion adequately in this work, but will be able to
address this more completely in future work using all five CANDELS fields. 

In the discussion of the violent channel, the enhanced SF in high \sersic\ AGN hosts could instead be taken as 
evidence for a starburst in these galaxies followed by rapid quenching within the last 100 Myr, 
the characteristic timescale over which a bolometric tracer such as the FIR retains memory of a burst of SF \citep{hayward14}.
Studies of the star-formation history of local emission-line selected AGNs in spheroidal hosts indicate that they underwent
a recent starburst \citep{schawinski10}, possibly related to the event responsible for their morphological transformation.
However, a scenario where gas brought in from external accretion onto already quenched spheroidal galaxies, which
then inspires both a starburst and AGN activity, can also explain these results \citep{simoes-lopez07,martini13}, 
so this is not a strong constraint on the extent of this process.

Nevertheless, if we assume that all clumpy, interacting and high \sersic\ AGNs at $0.5<z<1.0$ are triggered by a violent
channel, we can place a rough upper limit on the fraction of AGNs fuelled this way by adding the fractions of all three
categories of hosts. Clumpy hosts (CM$>0.5$) and interacting hosts (IM$>0.5$)
combined (including overlaps) account for $\approx 30$\% of the population, while including post-merger elliptical
hosts (with $n>3.5$) brings the fraction up to $\approx 60$\%. Thus, potentially, more than half of the population of low and moderate
luminosity AGNs at these redshifts may be fueled by gas brought to their centres by violent mechanisms. A K-S test
indicates that these AGNs do not have X-ray luminosities that are different from the rest, implying, to the
degree we can test with our small sample size, potentially violently triggered SMBHs have accretion rates that are
similar to those fueled by other processes. 

At higher redshifts ($z\sim2$), AGNs and inactive galaxies both show identical fractions of clumpy or interacting/merging systems,
while both AGNs and inactive galaxies have similar SFRs irrespective of structure. In particular, merging
AGN hosts have identical mean SFRs and FIR detection rates as the mass-matched control. These results point to a
lesser role for the violent channel in directly fueling AGN activity, since synchronisation between star-formation
and nuclear activity appears to be weak. Rather, the significantly higher gas surface densities in high redshift galaxies
\citep{tacconi10, daddi10a, tacconi13}, coupled with a faster secular evolution timescale \citep{genzel08}, can effectively
disconnect galaxy scale evolution from nuclear fueling processes. If AGN fueling at high redshifts is primarily modulated
by small scale processes, then how do we understand the higher \sersic, rounder light profiles for AGNs in such high redshift hosts?
If this is primarily due to widespread contamination of these profiles by reddened nuclear light, and the true host structures are indeed
disk-like and consistent with other massive galaxies, then a strong case can be made that secular fueling is the primary mode
at $z\sim2$ \citep{schawinski11, kocevski12, rosario12, hopkins13}. However, if there is a preference for AGNs to be found in bulgy
galaxies (as energetic arguments seem to support), then the violent mode responsible for the formation of such bulges 
may still retain a critical role in fueling higher redshift AGNs. Further clarification will come through the careful 
assessment of the true light profiles of these AGN hosts in up-coming CANDELS studies.

\acknowledgements

This work is based on observations taken by the CANDELS Multi-Cycle Treasury Program with the NASA/ESA HST, which is operated by the Association of Universities for Research in Astronomy, Inc., under NASA contract NAS5-26555. PACS has been developed by a consortium of institutes led by MPE  (Germany) and including UVIE (Austria); KUL, CSL, IMEC (Belgium); CEA, 
OAMP (France); MPIA (Germany); IFSI, OAP/AOT, OAA/CAISMI, LENS, SISSA 
(Italy); IAC (Spain). This development has been supported by the funding 
agencies BMVIT (Austria), ESA-PRODEX (Belgium), CEA/CNES (France),
DLR (Germany), ASI (Italy), and CICYT/MCYT (Spain). FEB acknowledges support from Basal-CATA PFB-06/2007, CONICYT-Chile (through FONDECYT 1101024, Gemini-CONICYT 32120003, "EMBIGGEN" Anillo ACT1101), and Project IC120009 "Millennium Institute of Astrophysics (MAS)", funded by the Iniciativa Cientifica Milenio del Ministerio de Economia, Fomento y Turismo. DMA acknowledges support from the Science and Technology Facilities Council (STFC) grant ST/I001573/1 (D.M.A.) and the Leverhulme Trust. We thank Victoria Bruce for helpful discussion.

\bibliographystyle{aa}

\bibliography{candels_morph_sfr}

\appendix

\section{The inclusion of a central point source in GALFIT fits}

Nuclear activity can systematically alter the appearance of AGN host galaxies by adding excess emission in the centre, which
concentrates the light profile, increases the resultant best-fit \sersic\ index and makes the system appear more circular.
Even at the highest attainable resolutions, the nucleus is completely unresolved in distant galaxies, appearing as a point source.
It is reasonable, therefore, to model AGN host galaxies as a combination of galaxy structural components and a central PSF. However,
bulges, which also make light profiles more central concentrated, cannot be easily distinguished from low levels of nuclear point
source emission. Galaxy bulges at $z\sim2$ have characteristic half-light diameters of $\sim2$ kpc or 0.24" \citep[e.g.][]{bruce12}, 
only slightly larger than the FWHM of the WFC3/F160W PSF (0.15"). Therefore, accurately distinguishing between bulges and
point sources requires careful modelling of the instrumental PSF, and even then may still be inaccurate, since galaxies do not
usually have regular light profiles. 

\begin{figure*}[h]
\includegraphics[width=\textwidth]{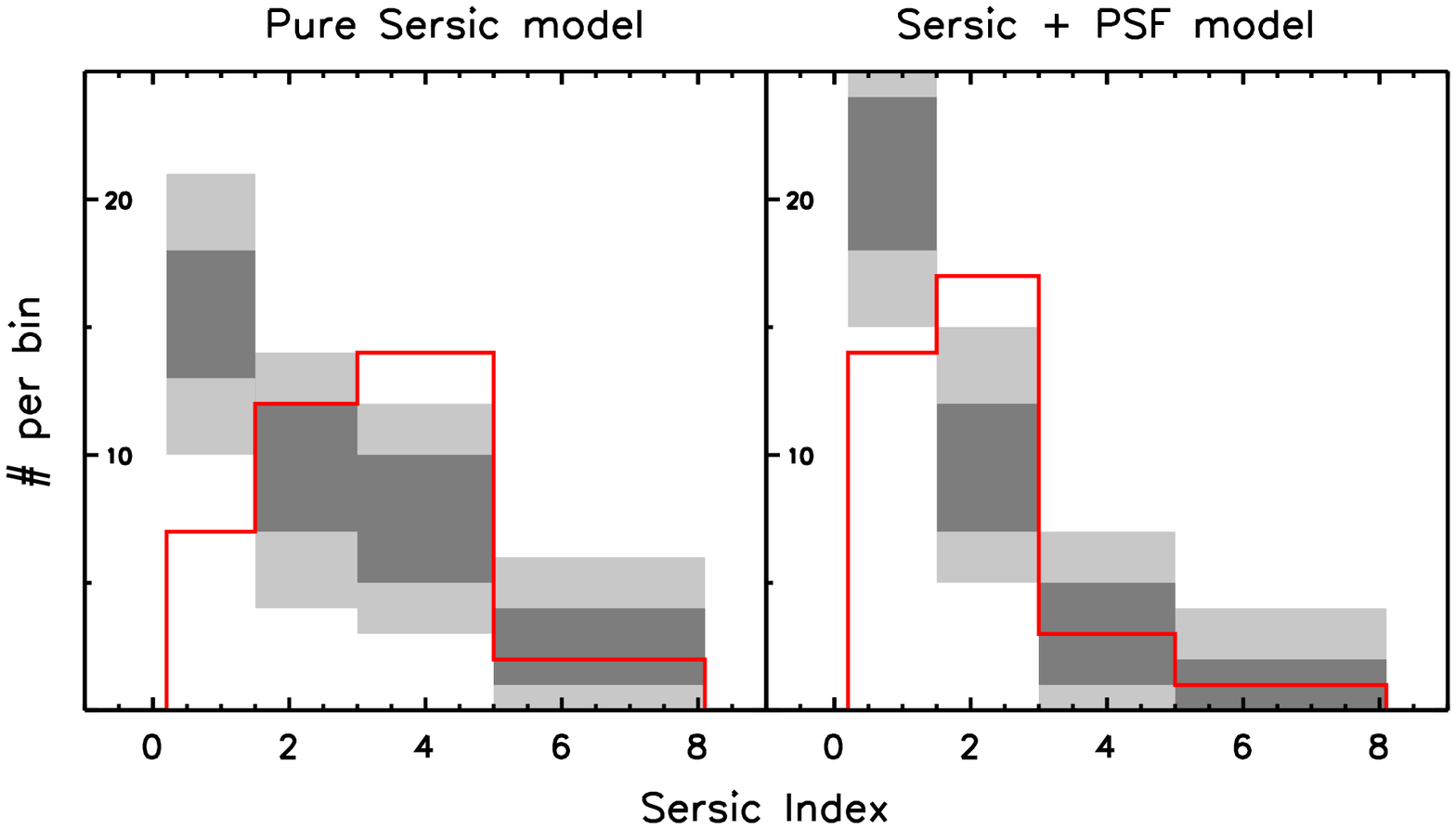}
\caption{Distributions of the \sersic\ index ($n$) from GALFIT fits to AGNs and inactive control galaxies
in the WFC3/F160W band. 
1000 draws of a mass-matched control sample are analysed to determine 
the $1\sigma$/$2\sigma$ scatter in the $n$ distributions for inactive galaxies, shown as dark/light
grey zones in the histograms. Red open histograms show the distributions for the AGNs.
Left: Results from fits of a single \sersic\ elliptical model. Right: Results from fits with two components - 
a \sersic\ elliptical model and a central point source.
The inclusion of a point source results in lower \sersic\ indices for both AGNs and inactive galaxies.}
\label{psf_test1}
\end{figure*}

To test the performance of GALFIT fits with a PSF component for real $z\sim2$ galaxies, we compared single \sersic\
and two component (\sersic\ + central PSF) fits to the F160W images of galaxies in the CANDELS GOODS-S field. 
This field has the deepest NIR imaging among the CANDELS fields. For this exercise, we used a subset of galaxies
from the full sample introduced in Section 2.4, which have been carefully fit with different light profile models 
for a study of the bulge properties of distant galaxies by \citep{lang14}.
Details of the methodology and fitting setup are published in that work. Only galaxies at $1.5<z<2.5$ were considered in
this exercise, since it is at these redshifts where the difference in light profiles between AGNs and inactive galaxies is most
pronounced (Section 3.3). The galaxies were all selected to have with \smass$>10^{10}$ \msun.

Each galaxy was first fit with a single elliptical
\sersic\ profile, iterated numerous times over a grid of initial values to prevent the best fit from falling into a local minimum. The best fitting
single \sersic\ profile was then used to initialise a second fit with the addition of a PSF component to the galaxy model.
The centre of the PSF was restricted to within 2 pixels (0.12") of the centre of the \sersic\ profile, and its flux was initialised to 1\% of 
the integrated magnitude of the galaxy. All galaxies, whether identified as inactive or active, were fit identically in this fashion.

In Figure \ref{psf_test1}, we compare the resultant \sersic\ index distributions of the X-ray AGNs to those of mass-matched inactive galaxies, where
the distributions for the inactive galaxies have been determined using the bootstrapping procedure discussed in Section 2.4.1. 
In the left panel, we show the results from single \sersic\ fits. These are qualitatively similar to those shown in Figures \ref{galfit_distributions}
and \ref{sersic_comps}, in that AGNs typically show significantly higher \sersic\ indices than inactive galaxies. In detail, we find a slightly
higher fraction of $n=4$ AGNs than in the full sample. However, given the small number of AGNs in this subsample, the difference
could arise from Poissonian variation. 

In the right panel, we plot the distributions of the best-fit \sersic\ index of the galaxy component in the two component fits. As expected, including
a PSF component has lowered the resultant $n$ of the AGN hosts, greatly increasing the fraction of disk-dominated systems. On the other hand,
an comparison of the distributions of the inactive galaxies between both panels in the Figure also demonstrates a reduction in the typical
$n$ for these galaxies as well. Since the inactive population is not expected to show widespread nuclear point source emission, we conclude
that simple two component fits as used in this exercise also tends to remove light from potential bulge components, systematically leading
to best-fit \sersic\ index distributions that are too disky.

\begin{figure}[h]
\includegraphics[width=\columnwidth]{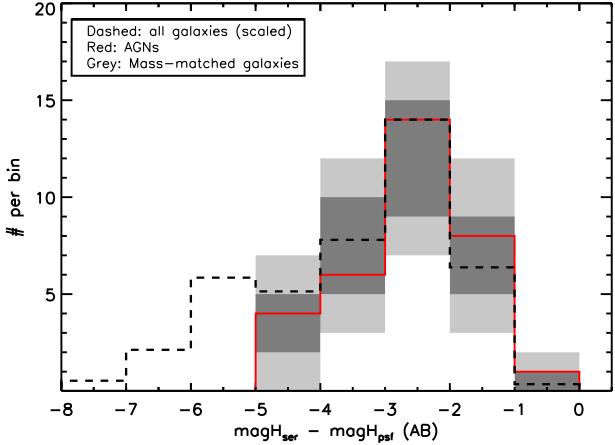}
\caption{Distributions of the flux ratio of the \sersic\ component to the PSF component (in magnitudes) 
from two component GALFIT fits to AGNs and inactive galaxies in the WFC3/F160W band. 
1000 draws of a mass-matched control sample are analysed to determine 
the $1\sigma$/$2\sigma$ scatter in the $n$ distributions for inactive galaxies, shown as dark/light
grey zones in the histograms. Red open histograms show the distributions for the AGNs. The dashed
black line is the distribution for the full inactive galaxy sample for which two component fits were performed,
including many lower mass galaxies not widely found among AGN hosts. This histogram has been scaled down to
overlap with the peak of the AGN histogram, to allow a simple visual comparison of the distributions.}
\label{psf_test2}
\end{figure}

Another valuable test of these fits is a comparison of the fraction of light in the PSF and galaxy components for AGNs and mass-matched inactive
galaxies. The nuclear luminosities of X-ray AGNs will be higher than any weak or heavily obscured nuclear emission that may remain undetected
in the inactive population. If the central excesses are indeed due to widespread nuclear point source contamination, we expect to find higher PSF fractions
among AGNs. In Figure \ref{psf_test2}, we plot histograms of the difference between the H-band magnitudes of the best-fit 
\sersic\ component and the best-fit PSF components from our fits. In addition to the AGNs (red) and mass-matched inactive galaxies (grey regions),
we also show the full distribution for galaxies with two-component fits (black dashed line), which includes many more low mass galaxies than generally
found among AGN hosts. The latter histogram has been scaled down in number to allow a visual comparison to the other distributions in the Figure.

Despite their higher nuclear luminosities, the two component fits yield essentially indistinguishable PSF fractions in the AGNs and equally massive
inactive galaxies. This is not simply due to limitations of the fits or local minima, since the overall distribution of PSF fractions includes a long tail to
very low values not found among the more massive AGN hosts (or other massive inactive galaxies). These two component fits were initialised 
with 5 mag between the PSF and the \sersic\ component, but the best-fit difference is about 2.5 mag (10\%). Considering that the excess central light
is found to a similar degree both in AGNs and inactive galaxies, one may conclude that the central excess is likely to arise in a bulge rather than
in a nuclear point source. This is consistent with arguments based on central colours and energetics from Section 3.3.1. 
We refrain from commenting on bulge fractions here - this requires detailed bulge+disk decomposition fits to both AGNs and inactive galaxies.

\section{Comparison with the CANDELS study of \citet{kocevski12}}

\citet{kocevski12} studied structural differences between AGN and mass-matched inactive control galaxies at $1.5<z<2.5$ in the CDF-S using essentially
the same visual classification scheme as in this work. However, there are important differences related to sample selections that should be
borne in mind when comparing our results to theirs. 

The X-ray source catalogs used in the two CANDELS studies are based on distinct reductions and source detection algorithms, leading
to different sized parent samples. 
We employ the CDF-S 4Msec catalog of \citet{xue11} which consists of 740 sources, while \citet{kocevski12} use a more conservative catalog 
of 569 sources. The differences between these catalogs
are discussed in \citet{rangel14}. In addition, we adopt AGN-specific photometric redshifts from \citet{luo10}, while \citet{kocevski12}
take redshifts from \citet{wuyts08} which are not optimised for AGNs. Despite these differences, the number of X-ray AGNs at $1.5<z<2.5$,
after the application of a lower \lhard\ limit, is nearly the same in both studies ($\approx 70$). Unlike us, however, \citet{kocevski12} do not
restrict their sample to the GOODS-MUSIC footprint or apply cuts in $m_{H}$ and \smass. Our sample of AGNs for the visual classification analysis
at $1.5<z<2.5$ is 55, in total after cuts, a reduction of 25\%. This smaller sample shares the same uniform analysis, quality checks and photometric
selections of the full GOODS-MUSIC dataset, enabling a consistent analysis of stellar mass and other galaxy properties between
AGNs and inactive galaxies in this work.

We can compare the fractions of disks and spheroids in our sample of AGNs with those reported by \citet{kocevski12}. From their Table 1, $\approx 80$\%
of AGNs are classified to have a visible disk or spheroid. In contrast, 100\% of our AGNs have one of these components. This is due to the
$m_H < 24.5$ cut applied to the galaxies in our visual classification catalog; we have less AGNs in our sample, 
but all have accurate structural assessments. As a consequence, our fractions of disks and spheroids will necessarily be higher than those
in \citet{kocevski12}, simply due to our different sample sizes. Therefore, we scale our fractions down by a factor of 1.25 to ease the
comparison.

In our AGN sample, we find disk galaxy fractions of $55_{-5}^{+5}$\% of which $13_{-3}^{+5}$\% are pure disks, with no reported spheroid component.
These fractions may be compared to $51_{-6}^{+6}$ and $17_{-4}^{+5}$\% respectively from Table 1 of \citet{kocevski12}.  We also find pure spheroid
fractions of $25_{-5}^{+5}$\% compared to $26_{-5}^{+6}$ from \citet{kocevski12}. As for morphologically disturbed
systems (Section 3.4), the fractions of visual disks and spheroids are completely consistent in both CANDELS works. Our
ability to reproduce the results of \citet{kocevski12}, despite the differences of approach and numbers of classifiers, 
highlights the stability of the CANDELS visual classification scheme. 

This being said, we prefer in this work to use an analytical measure of the galaxy light profile rather than visual measures of diskiness. 
As stated in Section 5.2, a visual classifier has difficulty discriminating between subtle variations in the light profile gradient. 
For e.g., inactive galaxies at $z\sim2$ that are classified visually as having dominant disk components exhibit a range in \sersic\ index of 
$0.7$--$3.2$ (80th percentile), while the range is $1.7$--$6.6$ for spheroid-dominated galaxies. 

\end{document}